%% file: paper.tex
\newif\ifarxiv
\arxivtrue
\newif\ifbbl
\bbltrue
\documentclass[conference]{styles/IEEEtran}

\ifarxiv

\pdfoutput=1 

\fi

\usepackage{cite}
\usepackage{balance}

\usepackage[pdftex]{graphicx}
\usepackage{grffile} 
\graphicspath{{figures/}}
\DeclareGraphicsExtensions{.pdf}

\usepackage[cmex10]{amsmath}
\usepackage{subfig}
\usepackage{wrapfig}

\usepackage{url}
\usepackage[usenames,dvipsnames]{color}

\usepackage[colorlinks=true, citecolor=OliveGreen, linkcolor=BrickRed, urlcolor=MidnightBlue, final, pdftex]{hyperref}

\usepackage{xspace}
\usepackage{algorithm}
\usepackage[noend]{algpseudocode}

\usepackage{comment}
\usepackage{fixltx2e} 

\newcommand{\etal}{\emph{et al.}}

\newcommand{\point}[1]{\noindent\textbf{#1}.}

\renewcommand{\S}{Section~}

\newcommand{\eg}{\emph{e.g.}}

\newcommand{\ie}{\emph{i.e.}}

\newcommand{\ps}{TAPS\xspace}
\newcommand{\compactify}{\settowidth{\labelsep}{o} \settowidth{\labelwidth}{o} \settowidth{\labelindent}{o}}

\title{Avoiding The Man on the Wire: Improving Tor's Security with Trust-Aware Path Selection}
\author{
\IEEEauthorblockN{Aaron Johnson\IEEEauthorrefmark{1},
Rob Jansen\IEEEauthorrefmark{1},
Aaron D. Jaggard\IEEEauthorrefmark{1},
Joan Feigenbaum\IEEEauthorrefmark{2} and
Paul Syverson\IEEEauthorrefmark{1}}
\IEEEauthorblockA{\IEEEauthorrefmark{1}U.S. Naval Research Laboratory, \{aaron.m.johnson, rob.g.jansen, aaron.jaggard, paul.syverson\}@nrl.navy.mil}
\IEEEauthorblockA{\IEEEauthorrefmark{2}Yale University, joan.feigenbaum@yale.edu}
}

\ifarxiv
\else

\IEEEoverridecommandlockouts
\makeatletter\def\@IEEEpubidpullup{9\baselineskip}\makeatother
\IEEEpubid{\parbox{\columnwidth}{This paper is authored by an employee(s) of the United States
    Government and is in the public domain. Non-exclusive copying or redistribution is allowed,
    provided that the article citation is given and the authors and agency are clearly identified
    as its source.  \\
    NDSS '17, 26 February - 1 March 2017, San Diego, CA, USA\\
    Internet Society, ISBN 1-1891562-46-0\\
    http://dx.doi.org/10.14722/ndss.2017.23307
}
\hspace{\columnsep}\makebox[\columnwidth]{}}

\fi

\begin{document}

\maketitle

\input{sections/abstract}
\input{sections/introduction}
\input{sections/attacks}
\input{sections/model}
\input{sections/metrics}
\input{sections/algorithm}
\input{sections/experiments}
\input{sections/performance}
\input{sections/errors}

\input{sections/propagation}
\input{sections/related}
\input{sections/conclusion}

\point{Acknowledgments}
We thank Ryan Wails for contributing to the results in Section~\ref{sec:attacks}.
The work at NRL was supported by ONR.
Joan Feigenbaum's research was supported in part by NSF grants CNS-1407454 and CNS-1409599, DHS
contract FA8750-16-2-0034, and William and Flora Hewlett Foundation grant 2016-3834.

 
\newcommand{\BIBdecl}{\setlength{\itemsep}{0\baselineskip plus 0.1\baselineskip minus 0.1\baselineskip}}
\balance
{\footnotesize

\ifbbl 
\input{paper.bbl}

\else 
\bibliographystyle{styles/IEEEtranS}
\bibliography{references}
\fi

}

 

\end{document}

%% file: sections/abstract.tex
\begin{abstract}
Tor users are vulnerable to deanonymization by an adversary that can observe some Tor relays
or some parts of the network. We demonstrate that previous network-aware path-selection
algorithms that propose to solve this problem are vulnerable to attacks across multiple Tor
connections. We suggest that users use trust to choose the paths through Tor that
are less likely to be observed, where trust is flexibly modeled as a probability
distribution on the location of the user's adversaries, and we present the Trust-Aware Path
Selection algorithm for Tor that helps users
avoid traffic-analysis attacks while still choosing paths that could have
been selected by many other users. We evaluate this algorithm in two settings using a high-level map
of Internet routing: (\emph{i}) users try to avoid a single global
adversary that has an independent chance to control each Autonomous System organization, Internet 
Exchange Point organization, and Tor relay family, and (\emph{ii}) users try to avoid 
deanonymization by any single country.  We also examine the performance of Trust-Aware Path
selection using the Shadow network simulator.
\end{abstract}

%% file: sections/introduction.tex
\section{Introduction}

Tor is a popular tool for low-latency anonymous communication, with over an estimated
1.5 million daily users. In order to use Tor to communicate with others, clients
choose a three-hop path from the set of over 7000 relays volunteering
bandwidth to the network. In order to balance load among the relays, and in particular
to optimize the
allocation of Tor's limited bandwidth, the default path selection
algorithm is \textit{bandwidth-weighted} so that a client will select a relay
with a probability equivalent to the ratio of that relay's available bandwidth
to the total network bandwidth capacity. Clients communicate with arbitrary
Internet hosts via a cryptographic \emph{circuit} with a layer of encryption for each
of the three relays on its path, which are termed the \emph{entry guard},
\emph{middle}, and \emph{exit},
according to their position on the path. Because this circuit is built
using a telescoping process, it provides unlinkability against a passive,
non-global adversary that cannot observe both ends of the circuit.

Unfortunately for many Tor users, a global adversary that can observe both ends
has become a very real and significant
threat. An adversary in such a position can perform a ``first-last''
traffic-correlation attack for any of the circuit's \emph{streams} (\ie{}, TCP connections
to destinations multiplexed over circuits)
by using similarities in the volume and timing of the traffic at both ends to match them with
each other, thereby deanonymizing the user. These techniques are efficient and
effective~\cite{torta05}. In order to carry out traffic-correlation attacks, an
adversary must be in a position to (\emph{i}) observe Tor traffic on an Internet
path between a client and its chosen entry guard or control
that entry guard, and (\emph{ii}) observe an Internet path between the
selected exit relay
and the destination or control the exit or destination. Due to Tor's
volunteer-relay model and its bandwidth-weighted path-selection algorithm, an
adversary may get in a position to observe a large amount of traffic simply
by running a fast relay, and it can otherwise observe traffic by controlling or
coercing entities on the paths to and from relays including Internet Service
Providers (ISPs), Autonomous Systems (ASes), and Internet Exchange Points
(IXPs).

Previous approaches to improving resilience against traffic observation and
correlation attacks have been limited in nature and only consider specific
threats. One main approach focuses on defeating an adversary that observes an
AS or
IXP~\cite{feamster:wpes2004,tor-as,lastor,murdoch:pet2007,juen-masters,astoria-ndss2016,denasa-pets2016}
and suggests creating Tor circuits such
that the set of ASes and IXPs that appear on the Internet paths between the client and
guard is disjoint from the set of ASes and IXPs between the exit and destination.
However, this solution ignores the critical effects of multiple connections over Tor,
which under this approach leak increasing amounts of information that can allow the
adversary to determine a client's location.
We present attacks of this nature on Astoria~\cite{astoria-ndss2016}, a recent
proposal of this sort, with our results showing that even a moderately-resourced adversary
can identify the client's AS within seconds. These attacks have similar implications for
all path-selection proposals using this approach.

The other main approach focuses on an adversary that
can observe some Tor relays~\cite{trusted-set,jsdm11ccs} and suggests that,
for the most sensitive circuit positions, users
choose a small number of relays from among those the user trusts the most. However, this
approach leaks information to an adversary that can
eventually identify the trusted relays (\eg{}, via a congestion
attack~\cite{long-paths,howlow}) and uses a restrictive trust model.
No existing solution to traffic correlation attacks provides protection from the
variety of attacker resources and tactics that recent research and experience
suggests is realistic~\cite{ccs2013-usersrouted}.

In contrast, this paper develops defenses against traffic correlation that are based on a
general probabilistic model of network adversaries. Using this model we can
consider adversaries with diverse resources, such as those that observe network traffic
at any combination of
network providers, exchange points, physical cables (undersea or elsewhere), and
Tor relays. The model allows us to incorporate uncertainty about and randomness in an
adversary's actions. It also enables us to express common factors underlying
the adversaries' ability to compromise different network locations, such as shared legal
jurisdiction or corporate ownership. A user expresses a \emph{trust belief} about her
adversaries by specifying a probability distribution over their presence at a set of network
locations, and she turns this belief into a \emph{trust policy} by including a weight for each adversary indicating her relative level of concern about that adversary.

Using these trust policies, we design Trust-Aware Path Selection (TAPS), a novel
path-selection algorithm that uses trust in network elements to inform a
user's decision about how to select relays for its Tor path. Using \ps, clients
select paths so as to minimize the probability that an adversary is in a
position to observe both ends of their Tor circuit while at the same time
ensuring that their path selection behavior does not harm their security by
making them stand out from other clients. In addition to defending against a much
broader class of adversaries, \ps addresses the other deficiencies of prior proposals.
In particular, it provides security against an adversary that can monitor and link user
activity across multiple connections, influence how users make connections, and
identify the relays used repeatedly by a client.

In order to facilitate the adoption of \ps, we describe both a long-term and a short-term
deployment strategy. In the long-term strategy, all Tor users participate and use
the \emph{TrustAll} version of \ps to replace Tor's existing bandwidth-weighted algorithm.
In the short-term strategy, \ps provides the option for security-conscious users to use trust to
defend against traffic-correlation attacks while most users continue to use ``vanilla''
Tor (\ie{}, bandwidth-weighted path selection). We design the \emph{TrustOne} \ps version for
this case, in which users must both avoid traffic correlation and choose paths that blend in
with the vanilla Tor users.

We evaluate the security of \ps via simulation with modified versions of the Tor Path Simulator
(TorPS)~\cite{ccs2013-usersrouted}. This evaluation is done with respect to two plausible and
illustrative trust policies:
(\emph{i}) \textsf{The Man} policy, in which a single adversary has an independent probability of
compromising each \emph{AS organization} (\ie{}, group of ASes run by the same entity), IXP
organization, and self-declared \emph{relay family} (\ie{}, set of relays run by the same entity)
in Tor; and (\emph{ii}) the \textsf{Countries}
policy, in which each country is considered a potential adversary and observes all ASes, IXPs, and
relays located inside of it. Our analysis of \textsf{The Man}
policy for a popular Tor client location shows a reduction in the probability of a successful
first-last attack from about 0.7 to about 0.4 in TrustAll with typical web-browsing activity
over the first week in December 2013,
and from about 0.68 to as little as 0.1 in TrustOne with repeated connections over the same week
to a single IRC
server popular with Tor developers. Our analysis of the \textsf{Countries} policy over that week
shows a reduction in the median number of countries
that ``unnecessarily'' compromise a stream (\ie{}, compromise a stream when they don't contain both
the client and destination) from 5 to 2 in TrustAll with typical user behavior.

Our algorithms are designed not only to improve security, but also to
allow security to be traded off for performance. They achieve this by allowing clients
to configure the fraction of bandwidth weight that their trusted sets of candidate guards
and exits should exceed before making a bandwidth-weighted choice from them. This mechanism
results in a client selecting from among the most secure relays while still
making use of much of the network and doing so in a bandwidth-weighted manner. We explore these
trade-offs using the Shadow simulator~\cite{jansen2012shadow,shadowweb} and find that there exist
parameters that result in only a slight decrease in performance, and only for less than 5 percent of
users.

\ifarxiv

This is the full version of the conference paper~\cite{taps-ndss2017}. The additions in this version
include a discussion of repeated-connection attacks and a measurement-based study of the risk of
cross-circuit attacks.

\fi

%% file: sections/attacks.tex

\section{Attacks on Network-Aware Path Selection}\label{sec:attacks}
There have been several proposals to improve Tor security by considering the network entities
(\eg{}, AS or IXP) that can observe a
circuit~\cite{denasa-pets2016,astoria-ndss2016,lastor,juen-masters,tor-as,feamster:wpes2004}.
However, none of these works considers anonymity across multiple Tor connections. Any realistic
use of Tor involves multiple connections from the same user, though, and these connections are
linkable by the adversary in many important cases, which means that it doesn't suffice to
consider the security of an individual connection. Indeed, we present two specific attacks that
can deanonymize users by identifying and analyzing multiple connections from the same user.

While there have been several proposals for network-aware path selection, we focus on just one
for concreteness: Astoria~\cite{astoria-ndss2016}. It is among the most recent and complete
proposals, and it provides a valuable example as many of the others suffer the same weaknesses.
Astoria is designed to prevent deanonymization by an AS (or group of \emph{sibling} ASes controlled
by the same organization). To choose a circuit for a given destination, a client determines the
ASes between the client and its guards and between the exits and the destination. It then makes a
bandwidth-weighted random choice from among the guard-exit pairs such that no AS appears on both the
client side and the destination side. If no such pair exists, a linear program is used to determine
the selection distribution over guard-exit pairs that minimizes the maximum probability that some
AS appears on both sides. Circuits are reused for destinations within the same AS.


\textit{Chosen-Destination Attack}: Consider an adversary that
runs a number of web servers in different locations and runs some Tor relays. If a Tor user visits
one of the malicious web servers, the adversary can force the browser to visit the other
malicious servers and request resources that are linkable to the original request (\eg{}, images
with unique names). The adversary will then observe the pattern of exits chosen by the client to
visit the servers in different locations, and it will also observe some of the client's guards if
the malicious relays are ever selected in the middle position. This attack strategy applies more
generally to any situation in which the adversary can choose the destinations that the client visits
and is able to link those connections as originating at the same client.

Under this attack, path-selection algorithms that choose relays based on the client's location can
leak increasing amounts of information about that location with each additional destination visited.
This is the case for Astoria, and we demonstrate that this attack is effective on that system.

To demonstrate the attack, we construct an AS-level Internet map using traceroute-based topology
data and inferred AS relationships from CAIDA~\cite{caida} and BGP routing tables supplied by
Route Views~\cite{routeviews}. We use the data from these sources for October 2015. Routes between
AS locations on our map are inferred using the algorithm proposed by Qiu and Gao~\cite{Qiu05aspath},
which takes AS paths observed in BGP tables and extends them to other locations using shortest
``valley-free'' paths. Conversion of IPs to ASes is performed using routing prefix tables from
Route Views~\cite{routeviews-prefix2as}.

We also use data from Tor Metrics~\cite{tormetrics} to evaluate the attack's effectiveness as if it
were run on the Tor network in the recent past. We use the archived network consensuses and server
descriptors to determine such facts as which relays existed, what families they were grouped in,
what their relative bandwidth capacities were, and what their ``exit policies'' were. We use the
first network consensus from October 2015 and its listed descriptors.
We also use data from Juen~\cite{juen-masters} that identifies 414 Tor client ASes as ever being
observed making a Tor connection and shows their relative popularity. 389 of these appear in our
AS map and are used in our analysis.
The five most popular client ASes identified by Juen, in order from most popular, are 6128 (Cable
Vision Systems, US), 25019 (SaudiNet, SA), 8972 (PlusServer AG, DE), 6893 (Saitis Network, CH), and
15467 (Enternet Libercom, HU).

We simulate Astoria's path selections for 1000 destinations in ASes selected uniformly at random
from those 44626 ASes in a maximal fully-connected component in our map that also advertise
prefixes. For each destination AS, an IP is chosen arbitrarily from an advertised prefix, and a
connection to port 443 is simulated. We repeat the simulation 100 times. We suppose that the
adversary runs the 4 relays with
the largest probabilities of being selected as a middle, which have a total advertised bandwidth of
1.01 Gbps and a cumulative probability of 3.3\% of being selected as a middle.

Table~\ref{table:guard_obs_prob} shows how often the adversary observes some or all of the client's
guards when the client uses 3 guards. It shows that with 100 destinations the adversary observes all
guards with 30\% probability, and with 300 destinations it observes all guards with 94\%
probability. It sometimes observes guards even with no destinations due to being selected as
a guard.

\begin{table}[ht]
\centering
\vspace{-3mm{}}
\caption{\small Probability of observing the client's guards}
\label{table:guard_obs_prob}
\begin{tabular}{l|l|l|l|l|}
\cline{2-5} & \begin{tabular}[c]{@{}l@{}}Pr. 0 guards\\ observed\end{tabular} & \begin{tabular}[c]{@{}l@{}}Pr. 1 guard\\ observed\end{tabular} & \begin{tabular}[c]{@{}l@{}}Pr. 2 guards\\ observed\end{tabular} & \begin{tabular}[c]{@{}l@{}}Pr. 3 guards\\ observed\end{tabular} \\ \hline
\multicolumn{1}{|l|}{0 destinations}   & 0.96 & 0.04 & 0 & 0 \\ \hline
\multicolumn{1}{|l|}{100 destinations} & 0.04 & 0.14 & 0.52 & 0.30 \\ \hline
\multicolumn{1}{|l|}{200 destinations} & 0 & 0.01 & 0.25 & 0.74 \\ \hline
\multicolumn{1}{|l|}{300 destinations} & 0 & 0 & 0.06 & 0.94 \\ \hline
\end{tabular}
\end{table}

\begin{figure}[ht]
\centering
\includegraphics[width=1.0\columnwidth]{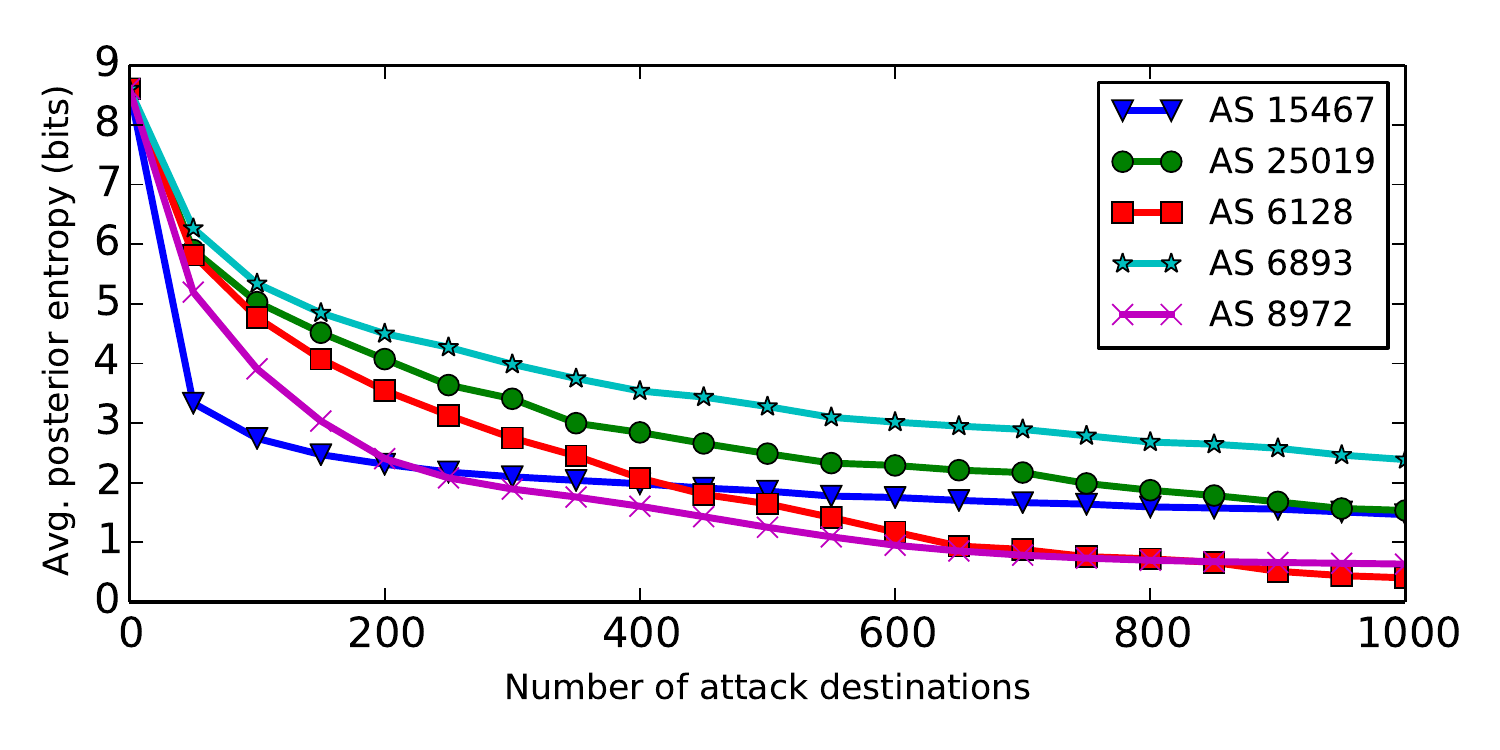}
\vspace{-4mm}
\caption{\small Avg. entropy of posterior client-AS distribution after chosen-destination attack on Astoria}
\label{fig:cda_entropy}
\vspace{-4mm}
\end{figure}

We then consider how well the adversary can guess the client's AS after the attack in the case that
he observes all of the client's guards. We again simulate Astoria's path selections for 1000
random destination ASes and repeat the simulation 100 times. We follow Nithyanand et
al.~\cite{astoria-ndss2016} in using 3 guards (1 and 2 guards yield similar results).
We suppose that the adversary uses a uniform prior distribution on the 389 client ASes. We then
compute the adversary's conditional distribution on the client ASes given the observed guard
set and the sequence of exits. The average entropy of the resulting distributions as we increase
the number of attack destinations is shown for the top 5 client ASes in
Figure~\ref{fig:cda_entropy}. It
shows an expected steady decrease in entropy for all client locations as the attacker uses more
destinations. By 300 destinations, all locations result in less than 4 bits of entropy on average,
and by 1000 destinations, the average entropy is less than 2.5 bits for all locations. Identifying
the client AS could be very dangerous for a Tor user, as it can identify the country of that user as
well as the ISP whose network logs could be used to completely identify the user. Note that this
attack could be completed within the time needed to construct circuits and open streams on them in
parallel, which is on the order of seconds.

\ifarxiv

\textit{Repeated Connections}: Consider a malicious destination to which a user repeatedly
connects over time. A new circuit is created for each connection that is separated by enough
time and inactivity (Tor uses a new circuit after 10 minutes by default, although Tor
Browser extends this for sufficiently-frequent connections). Suppose that the adversary can link
those connections as originating at the same user (\eg{}, by pseudonymous login at the
server). If the adversary also runs a relay, it can eventually learn both the client's guard set
(by being used occasionally as the middle relay) and the exit distribution used to reach the
destination from each guard (by observations at the destination). If these distributions
depend on the user's location, as in Astoria, the adversary can thus narrow down the user's
location over time.

\fi

\ifarxiv

\textit{Cross-Circuit Attack}: Consider a popular website that causes the user's browser to fetch
resources of (roughly) known sizes from a set of servers in other ASes. An AS-avoiding
path-selection algorithm like Astoria may create new circuits for these new destinations that place
a malicious AS between an exit and server on one connection and between the client and
a guard on another connection. If the adversary can link its
observations on those two connections, which it might by using the timing and amount of traffic on
the circuits, then it can deanonymize the user, effectively performing a \emph{cross-circuit}
correlation attack. Linking observations across circuits may be especially easy when the adversary
observes the first connection to a website from an exit because then it can compare traffic on
subsequent connections to a traffic fingerprint that it has previously measured for that website 
that contains the timing, amount, and destinations of traffic on connections initiated by that
website. In contrast to the chosen-destination and repeated-connections attacks, this attack can
be performed when the destinations are honest but some AS is malicious. This kind of attack enables
complete deanonymization by an AS via timing correlation even when each individual circuit is not
vulnerable to a correlation attack. Figure~\ref{fig:cross_attack} illustrates an example of the
cross-circuit attack.

\begin{figure}[ht]
\centering
\includegraphics[width=1.0\columnwidth]{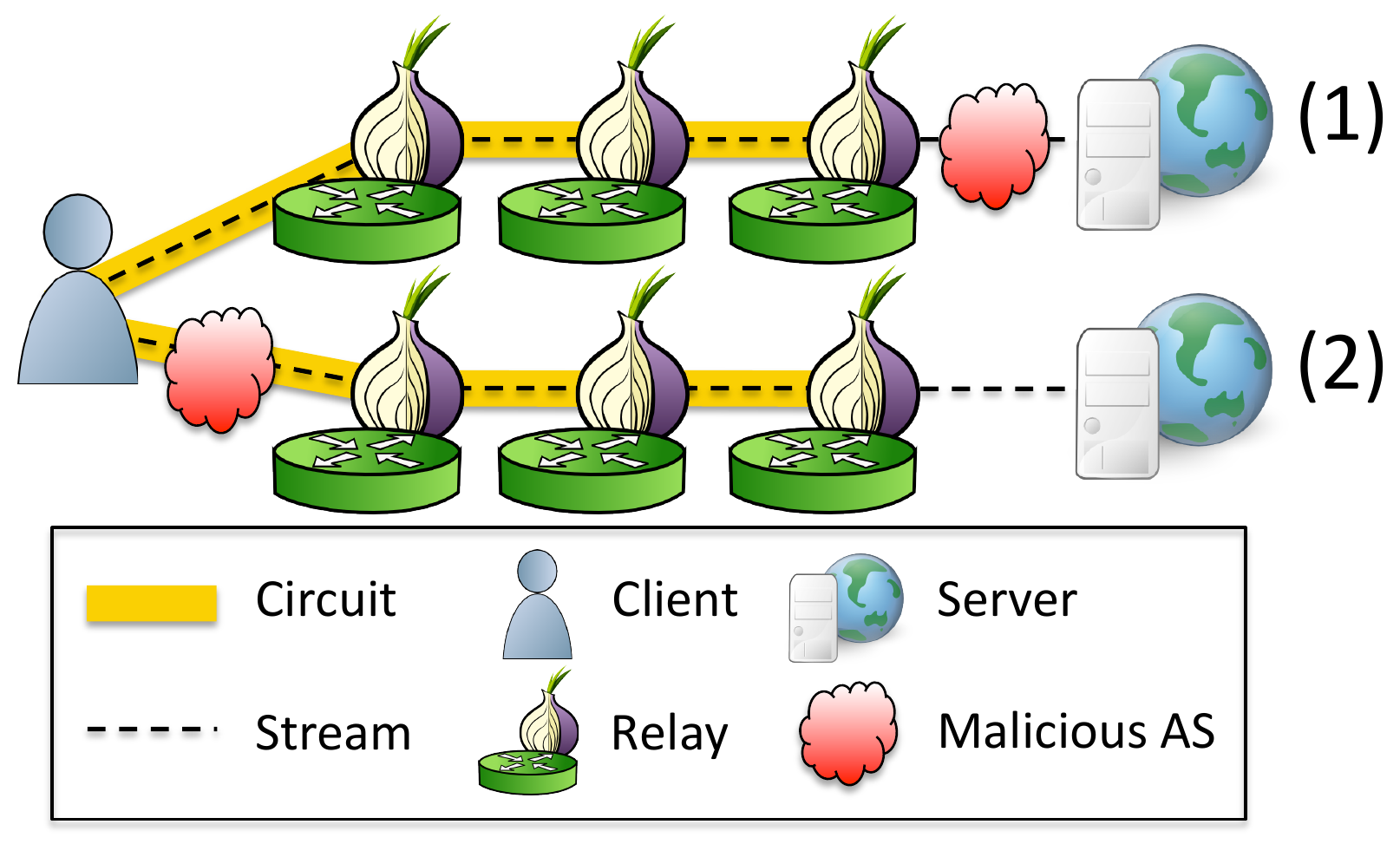}
\vspace{-4mm}
\caption{\small Example of a cross-circuit attack: a malicious AS observes the first connection made to a server during a website visit (1), then observes the client's connection to a guard on a subsequent connection during the same website visit (2), and then can conclude that the observed client visited the observed website.}
\label{fig:cross_attack}
\vspace{-4mm}
\end{figure}

We evaluate the potential vulnerability of Astoria to such an attack by examining how it constructs
and uses circuits when visiting popular websites. To measure the connections made during
such a visit, we use Tor Browser to fetch the front page from each of the Alexa global top
5000 websites~\cite{alexa}. This process is automated using
tor-browser-selenium~\cite{tor-browser-selenium}. We disable automatic updating and browser
caching, and we reload Tor Browser between each website visit. These steps help ensure that we only
consider streams initiated by the website visit and that we observe all streams normally
initiated by a fresh visit. We record all of the streams constructed during each visit using Tor's
control port~\cite{tor-control-spec}. For each visit, we extract a list of the IP addresses and TCP
ports to which streams are opened, excluding any streams generated by automated Tor events such as
consensus updates. We remove from these lists those IP addresses for which a mapping to AS number
doesn't exist, again using the Route Views data~\cite{routeviews-prefix2as} to determine the
mapping. We also manually inspect the returned page contents to verify that the website loaded
successfully.

We used Tor Browser v.6.0.4 to perform a crawl of the top 5000 sites on September 1st, 2016.
Of these, 2573 visits reached the desired site and resulted in a non-empty list of IP addresses and
ports. Reasons for failure include network error, interception by CloudFlare, a
redirect loop, timeout (set at 45s), and missing IP-to-AS mappings. Among these 2573 websites, the
number of unique pairs of IP address and TCP port connected to during the visit ranges from a
minimum of 1 to a maximum of 160 with a median number of 17. This suggests that there are many
popular websites for which Astoria may construct many new circuits and thus be vulnerable to a
cross-circuit attack.

To evaluate this, we simulate circuits in Astoria when visiting each website much
as we simulated Astoria when evaluating the chosen-destination attack. We again determine AS-level
paths using Qiu-Gao inference on our AS map, we use the same 389 popular client locations, and we
use the same data from Tor Metrics to model the relays in the Tor network. Then,
for each of the 2573 websites reached successfully and for each client location, we simulate
100 times how Astoria would create and use circuits to reach the sequence of addresses and ports
that were observed for the given website. We simulate Astoria with 3 guards, as this is the
recommended configuration. Fewer guards would reduce the possibility of a cross-circuit attack
(it would be eliminated with 1 guard) but would only reduce Astoria's protection against the
standard ``direct'' correlation attack that it is designed to mitigate.

The resulting number of circuits constructed varied significantly depending on the website.
If we aggregate the simulations by website and consider, for each
website, the median number of circuits constructed over all simulated visits to that website, we see
over all websites a minimum of 1 circuit, a median of 8, and a maximum of 54. If we instead
aggregate the simulations by client location and consider, for each client location, the median
number of circuits
constructed over all simulated visits by that client location, that value is 8 circuits for
\emph{every} client location. Thus the number of circuits constructed is highly dependent on the
destination. This result is unsurprising, as new circuits are constructed either when an IP address
in a new AS is encountered or when a new TCP port is incompatible with the existing exit's policy,
both of which are heavily dependent on the website. These results about circuit construction also
show that, for many popular websites, Astoria frequently provides many new circuit creations and
thus opportunities for a cross-circuit attack.

We evaluate the vulnerability of each simulated website visit to a direct correlation attack and to
a cross-circuit correlation attack. A visit is considered vulnerable to a direct attack
if there exists some AS $A$ such that, for some stream, $A$ appears both between the exit relay and
the destination host and between the client and guard. For this evaluation, we consider an AS to be
\emph{between} two Internet locations $B$ and $C$ if it is present on either the AS path from $B$ to
$C$ or the AS path from $C$ to $B$ (note that both $B$ and $C$ are thus considered between $B$ and
$C$). A website visit is considered vulnerable to a cross-circuit attack if there exists an AS $A$
such that (\emph{i}) $A$ is between the exit and the destination host on the \emph{first} stream,
(\emph{ii}) $A$ is between the client and the guard on some stream, and (\emph{iii}) $A$ is not
in a position to perform a direct attack during the visit as defined previously.

Figure~\ref{fig:cross_attack_results} shows how often Astoria is vulnerable to these attacks as a
distribution over all pairs of the 389 client locations and 2573 websites.
\begin{figure}[ht]
\centering
\includegraphics[width=1.0\columnwidth]{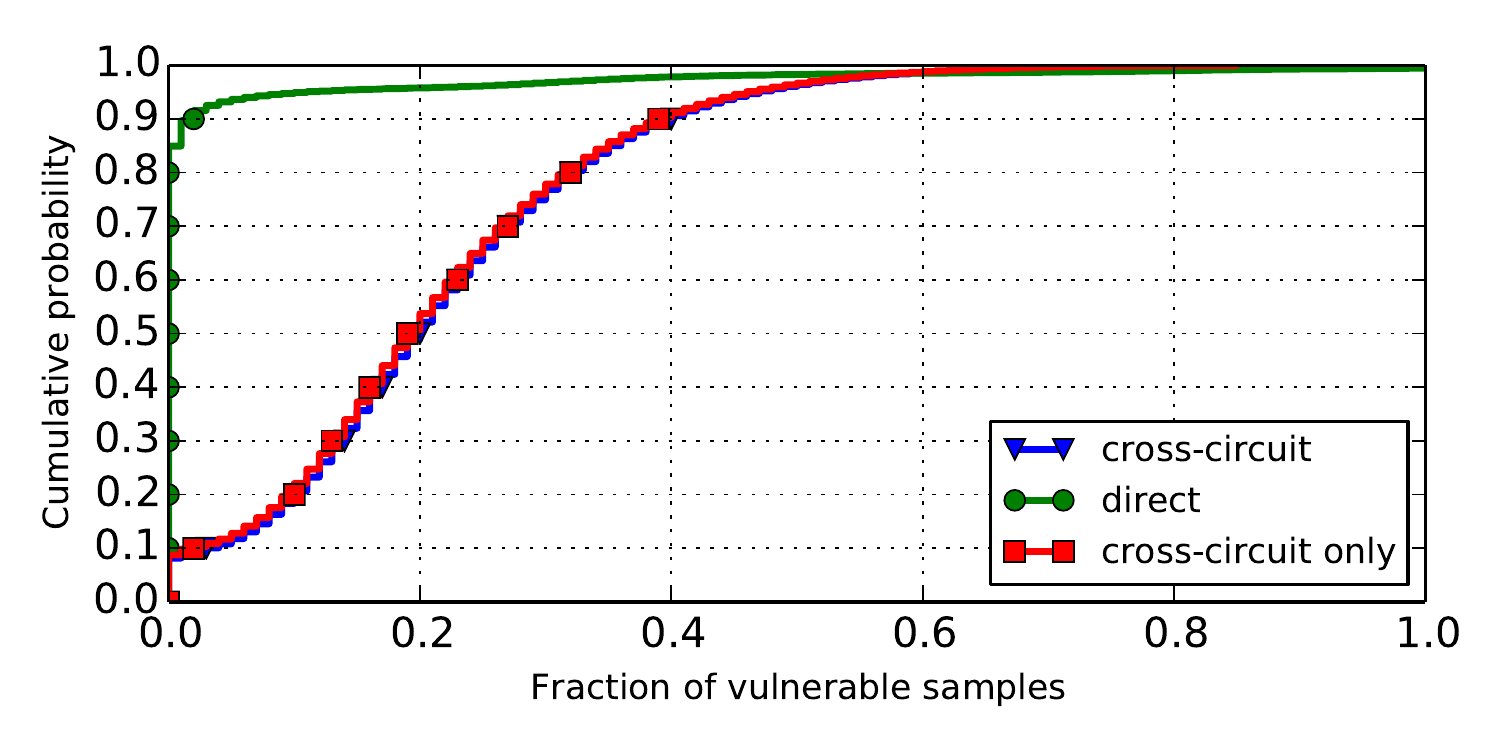}
\vspace{-4mm}
\caption{\small CDF over pairs of client location and website of the fraction of simulated
visits under Astoria that are vulnerable to a cross-circuit attack, a direct attack, and a
cross-circuit attack but no direct attack.}
\label{fig:cross_attack_results}
\vspace{-4mm}
\end{figure}
It shows that, while Astoria does effectively prevent most direct correlation attacks, for many
popular client locations and websites, users are frequently vulnerable to deanonymization via a
cross-circuit attack. Indeed, while the median fraction of simulated visits vulnerable to a direct
attack is only 0.03, the median fraction vulnerable to a cross-circuit attack is 0.2. To more
clearly consider the risk \emph{added} by the cross-circuit attack to the risk of direct attack, we
show the frequency distribution of visits that are vulnerable to a cross-circuit attack but not a
direct attack. The median frequency of this cross-circuit-only vulnerability is 0.19, which shows
that the cross-circuit attack is a major added risk.

These results expose a fundamental problem
with Astoria's strategy of varying the guard depending on the destination. Choosing two
different guards for two different destinations is
useful when for both guards some AS between the client and guard appears between the exits and
destination for one destination but not the other, which is precisely the situation when some AS is
in a position to perform the cross-circuit attack. This suggests that, unless the client can
somehow identify which pairs of connections are vulnerable to being linked as originating at the
same client, it should use few guards and choose them for circuits oblivious to the destination.

\fi

%% file: sections/model.tex

\section{Trust Model}\label{sec:model}
\subsection{Trust Policies}
We use the definition of network trust given by Jaggard \etal~\cite{trustrep-popets14}.
A trust belief is a probability distribution that indicates how likely adversaries are
to observe traffic at different network locations. A location is considered to be more trusted the
less likely it is that adversaries are observing it. Although the belief may be
expressed as the adversaries' success in compromising any number of relevant factors, such as relay 
software or physical location, ultimately
it must describe the probability that the adversaries observe traffic on the \emph{virtual links}
into and out of the Tor network. A virtual link is an unordered pair of network hosts, and the
\emph{entry virtual links} consist of client-guard pairs while the \emph{exit virtual links}
consist of destination-exit
pairs. An adversary is considered to observe a virtual link if it can observe traffic in
at least one direction between the two hosts. Although the representation of an arbitrary
distribution over all virtual links can be very large, Jaggard \etal~\cite{trustrep-popets14}
describe how distributions of likely Tor adversaries can be represented efficiently by aggregating
host locations (\eg{}, at the AS level) and by identifying a small set of relevant compromise
factors and indicating their dependencies in a Bayesian network.

To indicate how to trade off vulnerability to different adversaries, each user adopts a trust
policy that pairs her trust belief with a weight for each adversary. Each weight is a number in
$[0, 1]$ that indicates the relative level of concern that the user has for the associated
adversary. In this work, we assume that distinct adversaries do not collude.  If a user were worried
about two  adversaries colluding, she could combine her beliefs about them into those for a single
adversary. 

Trust policies are quite general and can easily be used with many kinds of beliefs
and sources of trust information. For example, previous work that considered each relay to have
an independent and individual probability of compromise~\cite{jsdm11ccs,trusted-set} can be
represented as a trust policy by including a single adversary and allowing him to compromise
each relay independently with its given probability. As
another example, previous work that considered as a potential threat each AS and
IXP~\cite{feamster:wpes2004,tor-as,juen-masters,murdoch:pet2007} can be represented as a trust
policy by including each AS and IXP as an adversary with equal weight and allowing each AS and IXP
to compromise with probability 1 all virtual links passing through it. Moreover, as described
by Jaggard \etal~\cite{trustrep-popets14}, trust policies can incorporate beliefs about a variety
of other sources of network compromise, such as software vulnerabilities, physical cable tapping,
and geographic location.

We do not expect that most individual users will craft their own trust policies. Indeed, doing so
is likely best left to experts unless the user has strong and idiosyncratic beliefs or concerns.
Rather, we envision that knowledgeable specialists, such as security researchers and professionals,
will provide
opinions about vulnerability to specific kinds of adversaries, and that institutions, such as
governments and consumer advocacy groups, will incorporate these opinions into trust policies that
are appropriate for their communities. An important special case of this is that we expect that the
Tor Project would select a \emph{default} policy that is in the broad interest of all Tor users, and
then would configure the standard Tor client to use it as well as provide any necessary supporting
data through the Tor network, much as Tor \emph{consensuses} (\ie{}, hourly documents describing
available relays) are distributed today by a set of directory authorities.

We will consider two specific trust policies in our analysis of \ps: (\emph{i}) \textsf{The Man},
which models a single powerful global adversary whose exact location isn't known with certainty,
and (\emph{ii}) \textsf{Countries}, which models each country as an adversary whose
locations are known exactly. Either of these policies constitutes a plausible default policy as well
as that of a particular user community. We now describe these models.

\subsection{\textsf{The Man}}
\textsf{The Man} represents a powerful adversary who may create, compromise, or coerce
the diverse entities that make up the Tor network. Specifically, we give \textsf{The Man}
an independent probability to observe each Tor relay family, AS organization, and IXP
organization. A relay self-identifies its family in a \emph{descriptor}~\cite{dir-spec} that it
uploads to the directory authorities, and
an AS or IXP organization is identified using public information as being controlled by the same
corporate or legal entity~\cite{Cai12b,ccs2013-usersrouted}.
Without any basis for differentiation, \textsf{The Man} compromises each AS and IXP organization
independently with probability 0.1. For relays, we consider that trust
may increase the longer a given relay has been active. This will not
guarantee protection against an adversary that is willing to
contribute persistently-high levels of service to the Tor network. However,
it can require adversaries to either make their own persistent
commitments to the network or to have compromised others who
have done so (and are thus most committed, experienced, and difficult to attack).
For \textsf{The Man}, we therefore assume each family
is compromised by the adversary independently with probability between
0.02 and 0.1, where the probability increases as the family's
longevity in Tor decreases. We calculate
longevity as follows: First, relay uptimes are calculated as the exponentially-weighted moving
average of the relay's presence in a consensus with Running, Fast, and Valid
flags with a half-life of 30 days.  A relay family's uptime is simply the sum
of its relays' uptimes. The probability that a family is compromised
is then taken as $(0.1-0.02)/(\mbox{family\_uptime}+1) + 0.02$.

\subsection{\textsf{Countries}}
As an alternative to \textsf{The Man}, the \textsf{Countries} trust policy includes as an
adversary each individual country in the world. A particular country
adversary compromises with probability 1 every AS or IXP that is located in that country
and no others. All country adversaries are given a weight of 1. This policy
illustrates a geographic perspective for Tor security, and it also demonstrates
how we handle multiple adversaries.

%% file: sections/metrics.tex

\section{Security Model and Metrics}\label{sec:metrics}


\subsection{Adversary Model}
As we have described in our trust model, we are considering an adversary who may
control or observe some Tor relays and parts of the Internet infrastructure. Note that
an important special case of this is that the adversary might observe the destination itself. From
these positions, we then analyze the adversary's success in deanonymizing users via the following
methods: (\emph{i}) performing a first-last correlation attack, (\emph{ii})
identifying the relays used on an observed connection, and (\emph{iii}) observing Tor
connections over time and linking them as belonging to the same user.

As described earlier, first-last correlation attacks are possible whenever the adversary
is in a position to observe traffic between the client and entry guard as well as between
the destination and exit. In such a situation, we assume that the adversary can immediately
determine that the observed traffic is part of the same Tor circuit and thereby
link the client with its destination.

Even when the adversary is not in a position to perform a first-last correlation attack, he still
may observe different parts of the circuit and use traffic correlation to link together those
parts. In such a case, if the observed relays on the circuit are unusually likely for
a particular client to haven chosen (\eg{}, because of atypical trust beliefs), then the adversary
may be able to identify the client even without direct observation. This is even more of a concern
if the adversary applies congestion or throughput attacks~\cite{long-paths,mittal-stealthy} to
\emph{indirectly} identify the relays on a target circuit. Therefore, we will consider the ability
of the adversary to identify the source and destination of an observed circuit based on knowledge
of its relays.

Finally, it is important to consider multiple connections over time instead of just one in
isolation. Every circuit that a client creates may give the adversary another opportunity obtain a
sensitive position or may leak more information about the client. This problem is made worse by
the fact that the adversary may be able to determine when some circuits are created by the same
client. This could happen, for example, if the adversary repeatedly observes traffic to the
destination and the client interacts with the same online service using
a pseudonym or requests a sequence of hyperlinked documents. Observe that in both of
these examples the linking is done using similarities in the content of traffic and not via any
weakness in the Tor protocol. Thus we will consider an adversary who can link
observed connections by client.

Note that we are not considering an adversary that can identify traffic content based only on
the timing and volume of data, that is, an adversary that can perform
\emph{website fingerprinting}~\cite{fingerprinting-wang-wpes13}. Also,
in reality we can expect \emph{adaptive} adversaries who continually learn and shift the
allocations of their resources, but we only analyze static adversaries in this paper. However,
adaptiveness can be captured to a certain extent already by defining trust policies with respect
to adversary behavior over time. That is, the compromise probability for a relay or virtual link
can represent the probability that it will \emph{at some point} during a given period be observed
by the adversary.

\subsection{Anonymity Metrics}
We evaluate anonymity using two kinds of metrics. The first kind will give empirical estimates of
the speed and frequency of first-last correlation attacks. The second kind will
provide worst-case estimates for the adversary's ability to identify the source or destination of
streams that are only partially observed.

First-last correlation attacks are relatively simple and result in complete deanonymization, and
therefore we are interested in accurate estimates for how likely they are to occur. Moreover,
their success depends on the behavior of the user and of the circuit-creation algorithm, and
therefore we can measure it empirically via simulation. Following Johnson et
al.~\cite{ccs2013-usersrouted}, we use the following as metrics: (i) The probability distribution of time until a client is deanonymized via a correlation attack; and (ii) The probability distribution of the fraction of streams that are deanonymized via a correlation attack.

Measuring the anonymity of multiple connections that are only partially observed is more difficult
because it isn't clear how successful the adversary can be at both linking separate streams and
indirectly identifying relays on a circuit. Therefore, we take a worst-case approach and consider
the adversary's ability to guess the source (resp. destination) of a sequence of streams that
have been linked as coming from the same client (resp. going to the same destination) and for
which the circuit relays have been identified. We measure this ability as the posterior
distribution over network locations. Note that we do not take into account the fact that the
adversary knows that streams for which the client or destination is unknown can only travel over
virtual links that the adversary does \emph{not} observe. Ruling out certain virtual links is more
challenging than just positively identifying traffic on a virtual link because it requires the
false negative rate for traffic correlation to be extremely low (in addition to the existing
requirement that the false positive be extremely low). Thus, we leave this extension to
our analysis to future work.


%% file: sections/algorithm.tex
\section{Trust-aware Path Selection}\label{sec:alg}
\subsection{Overview}
We describe two variants of the Trust-Aware Path Selection algorithms (\ps): (\emph{i}) TrustAll,
which is intended for system-wide deployment, and (\emph{ii}) TrustOne, which works with Tor's
existing path-selection algorithm and is intended for use by a minority of users.
Two \ps variants are needed to blend in with two different types of other users: those who do use
trust in path selection and those who do not. The main approach of
both algorithms is to choose guards and exits to avoid first-last correlation attacks while
also blending in with other users. Parts of this approach are shared with some previously-proposed
path-selection security improvements~\cite{tor-as,jsdm11ccs,lastor}. However, \ps includes several 
novel features that improve the security and performance issues of these proposals. We highlight
those features before proceeding to describe the algorithms in detail.

First, \ps uses an API that encapsulates flexible trust policies. These trust policies support
adversaries from a very general class of probabilistic models. As previously described, this class
can represent features such as uncertainty, multiple adversaries, and adversarial control of diverse
network elements.

Second, \ps clusters client locations and (separately)
destination locations. Each location cluster has a representative, and all locations in the
cluster are treated as if they were the representative. The main purpose of this clustering is to
prevent leakage of location information over multiple connections that would occur if paths were
selected differently for each pair of client and destination location. Treating all members of the
cluster as if they were the representative maintains anonymity within the cluster.
A secondary benefit is reducing the amount of information
needed to represent trust policies by reducing the number of paths to and from guards and exits
that need be considered.

Third, \ps treats its set of entry guards \emph{collectively}, that is, in a way that provides
security when multiple circuits, potentially using different guards, are considered. \ps chooses
each additional guard in a way that minimizes the additional exposure of
its entry paths. Moreover, once a set of entry guards is chosen, \ps doesn't prefer one guard in the
set over another when creating a connection to a given destination. This prevents the cross-circuit
attack discussed in Sec.~\ref{sec:attacks}. It also makes \ps compatible with the Tor's
current default configuration of one guard, as it does not depend on being able to choose the best
guard among several for a given destination.

Fourth, \ps provides a configurable tradeoff between security and performance by
parameterizing how much relay-selection deviates from ideal load-balancing.
The Tor network is under heavy load relative to its capacity~\cite{tormetrics}, and latency is
dominated by queuing delay at the relays~\cite{jansen2014kist}. Thus good load balancing
is essential to maintaining Tor's performance, which is itself a crucial factor in Tor's success.

\subsection{Trust API}
The \ps algorithms work with the trust policies described in Sec.~\ref{sec:model} via
an Application Programming Interface (API). Jaggard \etal~\cite{trustrep-popets14} describe how to
represent such policies
with a Bayesian network. However, such a representation may not be the most efficient for the
computations needed during path selection. Therefore, we abstract those computations into an API,
and we describe how they can be efficiently implemented for \textsf{The Man} and \textsf{Countries}
policies in Sec.~\ref{subsec:algorithm:api_imp}. We assume that the API is implemented by the
creator of the trust policy.

Several API functions take as an argument a network \emph{location}. There are
several possible levels of granularity at which a network location may be defined in \ps, such as
the Autonomous-System level or the BGP-prefix level.
Using more fine-grained locations will result in more accurate predictions
about the adversaries' locations and thus improve security, but it will also result in increased
runtime for the \ps algorithms (and likely for the API functions as well).

We assume that API users can provide a \emph{locations} data structure that (\emph{i}) allows the
locations to be enumerated, (\emph{ii})
includes each location's popularity rank for Tor clients, (\emph{iii}) allows IP addresses to be
mapped to locations, and (\emph{iv}) includes \emph{size} of each location
(\eg, the number of IP addresses originated by an Autonomous System).

We also assume that API users can provide a \emph{relays} data structure that
(\emph{i}) allows relays to be enumerated by unique identity keys (\eg, as represented by
fingerprints~\cite{dir-spec}), (\emph{ii}) includes the data in each relay's consensus entry (\eg,
the status flags, weight, and IP address), and (\emph{iii}) includes the data in each relay's
descriptor (\eg, the exit policy).

The trust API functions are as follows:

\begin{enumerate}[\compactify]
\item \textsc{LocationDistance}($\textit{loc}_1$, $\textit{loc}_2$, \textit{relays},
    \textit{weights}): This
    function returns an abstract \textit{distance} between two locations that measures the
    dissimilarity of the adversaries that appear on the network paths between the locations and the
    relays. This distance is the expected
    sum over \textit{relays} weighted by \textit{weights} of the total weight of adversaries that
    appear on one of the virtual links between the relays and $\textit{loc}_1$ and $\textit{loc}_2$
    but not the other. This function turns the set of locations into a metric space.

\item \textsc{GuardSecurity}(\textit{client\_loc}, \textit{guards}): This
    function returns a security \textit{score} for the use of the given \textit{guards} as entry
    guards by a client in location \textit{client\_loc}. The score must be in $[0, 1]$, and it
    should represent the expected total weight of adversaries \emph{not} present on the paths
    between \textit{client\_loc} and \textit{guards}, normalized by the sum
    of all adversary weights. Thus a higher score indicates higher security.

\item \textsc{ExitSecurity}(\textit{client\_loc}, \textit{dst\_loc}, \textit{guard},
    \textit{exit}): This function returns a security score for the use of \textit{guard} and
    \textit{exit} by a client in location \textit{client\_loc} to connect to a destination in
    \textit{dst\_loc}. The score must be a value in $[0, 1]$, and it should
    represent the expected total weight of the adversaries that either are not present on the path
    between \textit{client\_loc} and \textit{guard} or are not present on the path between
    \textit{dst\_loc} and
    \textit{exit} (\ie{}, those not able to perform a correlation attack), normalized by the sum of
    all adversary weights. Thus, again, a higher score indicates higher security.
\end{enumerate}

\subsection{TrustAll}
TrustAll consists of two separate processes:

\begin{enumerate}[\compactify]
\item \textsc{Cluster}: This process is run by the trust-policy provider (\eg{}, by the Tor
  directory authorities for the default policy). It clusters client and destination
  locations and makes the results available to clients. To maintain the anonymity sets provided by
  the clusters, this process should execute infrequently (\eg{}, every 6 months) and only to reflect
  significant changes in the trust on entry and exit virtual links. It takes the \textit{locations}
  and \textit{relays} data structures as inputs and produces \textit{clusters} as output, which is a
  mapping from each location chosen as a cluster representative to the set of locations in its
  cluster.
  
\item \textsc{Connect}: This process is run by a Tor client. It runs every time a new connection
  is requested. It uses the output of the \textsc{Cluster} process, the state of the client (\eg{},
  current network consensus, current circuits, and client IP address), \textit{locations}, and
  \textit{relays}. It may create a new circuit or reuse an existing one.
\end{enumerate}
We now detail these processes.

\subsubsection{Cluster}
Network locations are clustered twice. One clustering will be applied to the destination's location
during path selection, and the other will be applied to the client's location.
The output of a clustering is a partition of network locations with a single member of each cluster
in the partition designated as that cluster's representative. Client and destination clusterings are
performed slightly differently because a single client is likely to visit many destinations.
Therefore, if we were to bias destination clusters, we could potentially reduce security for all
clients on at least one connection, but we can bias client clusters towards the most likely
locations and improve security for most clients.

\paragraph{Clustering destination locations}
Destinations are clustered with a k-medoids algorithm~\cite{park-medoid-voronoi-2009}, modified to
produce balanced-size clusters. Balance is needed for good anonymity, as each cluster is an
anonymity set. The medoids of the resulting clusters are used as representatives. 
The destination-clustering algorithm takes two parameters: (\emph{i}) \textit{num\_clusters},
the number of clusters to produce, and (\emph{ii}) \textit{max\_rounds}, the maximum number of
assignment rounds. The clustering is accomplished as follows:
\begin{enumerate}[\compactify]
	\item Choose as an initial cluster representative a location uniformly at random from
	\textit{locations}.

    \item Choose the remaining \textit{num\_clusters}$-1$ cluster representatives by
    iteratively choosing the location with the largest distance to the representatives
    already chosen (\ie{}, the maximin distance), with distances determined by
    \textsc{LocationDistance}$()$.

    \item Assign locations to cluster representatives by greedily assigning to the smallest
    cluster (in terms of the total size of its locations) at a given time the location closest to
    its representative, as measured by \textsc{LocationDistance}$()$.

    \item Recalculate the cluster representatives by determining the location in
    each cluster with the smallest average distance to each other member in the
    same cluster.

    \item Repeat from step (3) if any cluster representative changed and there have been fewer than
    \textit{max\_rounds} assignment rounds.

    \item Return both the clusters and their representatives.
\end{enumerate}


\paragraph{Clustering client locations}
Client clustering uses known popular client locations as cluster representatives and
then clusters the remaining locations in one round. The client-clustering algorithm takes as input
\textit{num\_clusters}, the number of clusters to produce. It creates
the set of cluster representatives by selecting the \textit{num\_clusters} most-popular client
locations. The remaining locations are assigned to clusters by greedily assigning to the smallest
cluster at any time as in the destination clustering.

\subsubsection{\textsc{Connect}}
The \textsc{Connect} process is invoked when a Tor client is requested to connect to a destination.
We assume for now that any needed DNS resolution has been performed, and \textsc{Connect} has been
given a destination IP address. Section~\ref{subsec:algorithm:disc} discusses how DNS resolution
might occur.

The essential mechanism that TrustAll uses to improve security is to compute security
\emph{scores} for relays in the guard and exit positions and then to only select the highest-scoring
relays for those positions. Guard security scores are determined with \textsc{GuardSecurity}$()$,
which takes into account any existing guards when choosing a new one and thus provides
security with respect to the entire guard set. Exit security scores are determined with
\textsc{ExitSecurity}$()$,
which takes into account the guard to be used for the circuit and thus can mitigate first-last
correlation attacks.

Given scores for relays in position $p\in \{g, e\}$ ($g$ for guard and $e$ for exit), the
\textsc{SecureRelays}$()$ function (Alg.~\ref{alg:secure_relays}) is used to determine
the \emph{secure} relays, that is, those relays with high-enough scores to be selected for a given
position. Note that in Alg.~\ref{alg:secure_relays} \textsc{ReverseSort}($X, Y$) sorts the entries
$x\in X$ by descending values $Y[x]$, and \textsc{Length}($X$) returns the number of entries in
$X$.
\begin{algorithm}
\caption{TrustAll secure relays to use for position $p$}
\label{alg:secure_relays}
\begin{algorithmic}[0]
\Function{SecureRelays}{$\alpha_p, scores, relays, weights$}
\State $R \gets $ \Call{ReverseSort}{$relays, scores$}
\State $s^* \gets scores[R[0]]$ \Comment{Maximum score}
\State $n \gets $ \Call{Length}{$relays$} \Comment{Number of relays}
\State $S \gets \emptyset$, $w \gets 0$, $i \gets 0$
\While{$(scores[R[i]] \ge s^* \alpha^{su}_p) \land$\\
    \hspace{\algorithmicindent}\hspace{\algorithmicindent}$(1-scores[R[i]] \le (1-s^*) \alpha^{sc}_p) \land$\\
    \hspace{\algorithmicindent}\hspace{\algorithmicindent}$(i < n)$}
    \Comment{Add all safe relays}
  \State $S \gets S \cup \{R[i]\}$
  \State $w\gets w + weights[R[i]]$
  \State $i \gets i + 1$
\EndWhile
\While{$(scores[R[i]] \ge s^* \alpha^{au}_p) \land$\\
    \hspace{\algorithmicindent}\hspace{\algorithmicindent}$(1-scores[R[i]] \le (1-s^*) \alpha^{ac}_p) \land (w < \alpha^w_p) \land$\\
    \hspace{\algorithmicindent}\hspace{\algorithmicindent}$(i < n)$}
    \Comment{Add acceptable relays}
  \State $S \gets S \cup \{R[i]\}$
  \State $w\gets w + weights[R[i]]$
  \State $i \gets i + 1$
\EndWhile
\State \textbf{return} $S$
\EndFunction
\end{algorithmic}
\end{algorithm}
\textsc{SecureRelays}$()$ identifies the highest score $s^*$. It adds to the set of secure relays
all \emph{safe} relays, that is, relays with scores very close to $s^*$. Then it considers the
\emph{acceptable} relays, that is, relays with scores close to $s^*$ but not close enough to make
them safe. Acceptable relays are added in descending order of score until the desired fraction of
the total bandwidth in position $p$ is reached. Distinguishing safe from acceptable relays enables
improved load balancing when there are many highly-trusted choices available.

The parameters defining these sets are given as the list
$\alpha_p = (\alpha^{su}_p, \alpha^{sc}_p, \alpha^{au}_p, \alpha^{ac}_p, \alpha^w_p)$.
$\alpha^{su}_p$ and $\alpha^{sc}_p$ are used just for safe relays, and $\alpha^{au}_p$ and
$\alpha^{ac}_p$ are used just for acceptable relays. Safe and acceptable relays are defined using
the same method, but acceptable relays use less restrictive parameters, with
$\alpha^{au}_p \le \alpha^{su}_p$ and $\alpha^{ac}_p \ge \alpha^{sc}_p$.

The ``uncompromised'' parameter
$\alpha^u\in \{\alpha^{su}_p, \alpha^{au}_p\}$ is used to include a relay only if it has a
security score $s$ such that $s\ge s^* \alpha^u$. It must be that $\alpha^u\le 1$, or no
relays would qualify. $\alpha^u$ is thus the required fraction of the maximum possible expected
weight of adversaries with
respect to whom the circuit position is considered uncompromised. One effect of this constraint is
that relays with completely untrusted paths will not be chosen if there is at least one other
option.

The ``compromised'' parameter $\alpha^c\in \{\alpha^{sc}_p, \alpha^{ac}_p\}$ is used to
include a relay only if it has a score $s$ such that $1-s \le (1-s^*) \alpha^c$.
It must be that $\alpha^c\ge 1$, or no relays would qualify.
$\alpha^c$ is thus a limit on the multiple of the minimum possible expected weight of adversaries to
whom the circuit position is considered compromised. An effect of this constraint is that if
relays with completely trusted paths are available, then no other options are considered.

$\alpha^w_p$ represents the desired minimum bandwidth fraction of relays in position $p$ for
the secure relays. It will be reached if the safe and acceptable relays together constitute at least
that fraction. The \textit{weights} argument to \textsc{SecureRelays}$()$ maps relays to their
positional bandwidth to determine if and when $\alpha^w_p$ is reached.

Let \textit{client\_rep} be the representative location for the client's cluster and
\textit{dst\_rep} be the representative location for destination's cluster. The \textsc{Connect}
process proceeds as follows:

If the number $\ell$ of selected and responsive guards is less than the number $k$ desired
(\eg{}, $k$ is the value of \textsf{NumEntryGuards}~\cite{dir-spec}), then $k-\ell$ new guards are
selected. Each guard is added by (\emph{i}) creating \textit{scores} where each potential guard $g$
has score
\textsc{GuardSecurity}$(\textit{client\_rep}, G\cup\{g\})$, with $G$ is the current set of guards;
(\emph{ii}) identifying as the set of secure guards
$S = \textsc{SecureRelays}(\alpha_g, \textit{scores}, P, \textit{g\_weights})$, where $\alpha_g$
contains the guard security parameters, $P$ contains all potential guards not currently selected,
and \textit{g\_weights} contains the relays' weights for the guard position; and (\emph{iii})
randomly selecting from $S$ with probability proportional to \textit{g\_weights} (\ie{} making a
bandwidth-weighted choice).

Consider the existing circuits in reverse order of the time a stream was most-recently
attached. If current circuit $c$ is too \textit{dirty}, that is, a stream was first attached too
long ago (Tor's default dirtiness threshold is 10 minutes), then proceed to the next circuit.
Otherwise, let
$g_c$ be the circuit's guard, let $\alpha_e$ contain the security parameters for exit
selection, let \textit{exits} contain all potential exits for the desired connection according to
the criteria Tor currently uses that don't depend on the guard (\eg{}, a compatible exit policy),
and let \textit{e\_weights} contain the relays' weights for the exit position. Let
\textit{scores} contain
the exit scores, with $\textit{scores}[e] =$ \textsc{ExitSecurity}(\textit{client\_rep},
\textit{dst\_rep}, $g_c$, $e$) for all $e\in \textit{exits}$. Compute the set of secure exit relays
$S = $ \textsc{SecureRelays}($\alpha_e$, \textit{scores}, \textit{exits}, \textit{e\_weights}).
If the circuit's exit $e_c$ is in $S$, then reuse the circuit. Otherwise, proceed to the next
circuit.

If no suitable circuit has been found and reused, let $c$ be the circuit among those that are
not too dirty that most recently had a stream attached, and let $g_c$ be its guard. Choose a
new exit $e$ by (\emph{i}) creating \textit{scores} where each $e\in\textit{exits}$ has score
\textsc{ExitSecurity}$(\textit{client\_rep}, \textit{dst\_rep}, g_c, e)$; (\emph{ii})
identifying as the set of secure exits
$S = \textsc{SecureRelays}(\alpha_e, \textit{scores}, \textit{exits}, \textit{e\_weights})$; and
(\emph{iii}) randomly selecting from $S$ with probability proportional to \textit{e\_weights}.
Reuse $c$ through its first two hops but ``splice'' $e$ onto the end after the middle relay.
This effectively operates as a new circuit, but the handshakes through the first two
hops are not repeated to reduce the latency of creating it.

If no circuit exists that is not too dirty, create a new circuit as follows: (\emph{i})
choose a guard $g$ uniformly at random from the $k$ selected and responsive guards, (\emph{ii})
choose an exit $e$ as described for the case that a new exit is being spliced onto an existing
circuit with $g$ as the guard,
and (\emph{iii})
choose a middle node as Tor currently does given $g$ and $e$ (\eg{}, bandwidth-weighted random
selection).  Note that, in contrast
to vanilla Tor path selection, the guard and exit are not explicitly prevented from being contained
in the same /16 subnet or relay family. Instead, the threat of entry and exit paths being observed
by the same relay family or network is incorporated into the trust policy, and vulnerable paths are
avoided by \ps.

\subsection{TrustOne}
TrustOne path selection is designed to be used when most users are not using \ps and instead are
using vanilla Tor path selection. Thus slightly different behavior is required in order to fully
blend in with the larger group. Also, if most users do not use trust, then more secure parameters
can be used without impacting performance much.

As with TrustAll, TrustOne consists of a \textsc{Cluster} process and a \textsc{Connect} process.
\textsc{Cluster} is performed in the same way as in TrustAll. The \textsc{Connect} process differs
from that of TrustAll in the following ways:

\begin{itemize}[\compactify]
\item \textsc{SecureRelays}$()$ doesn't use the notions of safe and acceptable relays. It simply
orders relays by their score and chooses the most secure up to the desired bandwidth fraction.
The TrustOne version of this function appears in Alg.~\ref{alg:trustone_secure_relays}. Note that
the performance parameter is a single value (\ie{}, $\alpha_p = \alpha^w_p$). TrustOne
doesn't use the concept of acceptable relays because it must allow exit relays to be chosen the same
they are in vanilla Tor path selection, which in TrustOne will happen when
$\alpha^w_e = 1$. Also, TrustOne can omit the distinction between safe and acceptable relays
because load balancing is less important when few users are using trust.

\item Given a guard, potential exits (\ie{}, the set over which scores are computed with
\textsc{ExitSecurity}$()$)
are chosen exactly as they are in vanilla Tor path selection,
including in particular the constraints preventing exits and guards from sharing a family or /16
subnet. This prevents a TrustOne user from being identified as using non-standard path selection
(\eg{}, by a middle relay).
\end{itemize}

\begin{algorithm}
\caption{TrustOne secure relays to use for position $p$}
\label{alg:trustone_secure_relays}
\begin{algorithmic}[0]
\Function{SecureRelays}{$\alpha^w_p, scores, relays, weights$}
\State $R \gets $ \Call{ReverseSort}{$relays, scores$}
\State $S \gets \emptyset$, $w \gets 0$, $i \gets 0$
\While{$w < \alpha^w_p$} \Comment{Add desired fraction of relays}
  \State $S \gets S \cup \{R[i]\}$
  \State $w\gets w + weights[R[i]]$
  \State $i \gets i + 1$
\EndWhile
\State \textbf{return} $S$
\EndFunction
\end{algorithmic}
\end{algorithm}

Note that a client can choose not to protect the fact the he is using TrustOne instead of vanilla
Tor by setting a desired exit bandwidth fraction of $\alpha^w_e < 1$. He may do this when he
doesn't believe that revealing his use of TrustOne will reveal his identity, and using a smaller
$\alpha^w_e$ will improve his security against a first-last correlation attack by restricting
his circuits to more-secure exits.

\subsection{Trust API implementations} \label{subsec:algorithm:api_imp}
The efficiency of the trust API depends on the type of trust policies
used. For example, a user with trust in individual relays may only need to store a single
trust value for each relay and perform simple arithmetic computations for the API functions, while a
user with trust in Autonomous Systems may need to store an entire Internet topology and perform
routing inference. In general, because the API functions return the expectation of values that can
easily be computed if the compromised relays and virtual links are unknown, they can be implemented
by repeatedly sampling the adversary distributions. Thus the API functions are compatible with the
Bayesian-network representation of Jaggard \etal~\cite{trustrep-popets14}. However,
policies should use implementations that are efficient for their specific features. 

For \textsf{The Man} policy, the trust API functions need access to data describing the relay
families, AS organizations, IXP organizations, and the virtual entry and exit links on which each AS
and IXP organization has a presence. The API functions can easily be implemented efficiently for
\textsf{The Man} because there is a single adversary whose presence on a virtual link depends on the
compromised status of \textit{network entities} (\ie{}, relay families, AS organizations, and IXP
organizations) that are each independently compromised. We implement the API functions as follows:

\begin{itemize}[\compactify]
\item \textsc{LocationDistance}$(\textit{loc}_1$, $\textit{loc}_2, \textit{relays},
    \textit{weights})$: To compute
    this, consider each $r\in \textit{relays}$. Let $E_1$ be the set of network entities that exist
    between $r$ and both $\textit{loc}_1$ and $\textit{loc}_2$, let $E_2$ be the set of network
    entities that exist only between $r$ and $\textit{loc}_1$, and let $E_3$ be the set of
    network entities that exist only between $r$ and $\textit{loc}_2$. Let $p_r$ be the probability
    that the adversary is present on
    one of the paths from $r$ to $loc_1$ and $loc_2$ but not the other. $p_r$ is simply the
    probability that (\emph{i}) no $e\in E_1$ is compromised and (\emph{ii}) either some $e\in E_2$
    is compromised and no $e\in E_3$ is compromised or vice versa. The distance is computed as the
    \textit{weights}-weighted sum of $p_r$ over $r\in \textit{relays}$.
    
\item \textsc{GuardSecurity}$(\textit{client\_loc}, \textit{guards})$: Let $E$ be the set of network
    entities between \textit{client\_loc} and the \textit{guards}. The security score computed as
    the product of the probabilities that each $e\in E$ is individually uncompromised.
    
\item \textsc{ExitSecurity}$(\textit{client\_loc}, \textit{dst\_loc}, \textit{guard},
    \textit{exit})$: Let $E_1$ be the set of network entities that exist both between
    \textit{client\_loc} and \textit{guard} and between \textit{dst\_loc} and \textit{exit}, let
    $E_2$ be the set of network entities that exist only between \textit{client\_loc} and
    \textit{guard}, and let $E_3$ be the set of network entities that exist only between
    \textit{dst\_loc} and \textit{exit}. The security score is the product of the probability that
    no $e\in E_1$ is compromised and that either no $e\in E_2$ is compromised or no $e\in E_3$ is
    compromised.
\end{itemize}

For the \textsf{Countries} policy, the trust API functions need the list of all countries as well
as a data structure mapping each relay to its country and each virtual link to the countries it
passes through.
\textsc{LocationDistance}$(\textit{loc}_1$, $\textit{loc}_2, \textit{relays}, \textit{weights})$
is just a weighted sum over $r\in \textit{relays}$ of the number of countries on which the virtual
links $\{\textit{loc}_1, r\}$ and $\{\textit{loc}_2, r\}$ disagree.
\textsc{GuardSecurity}$(\textit{client\_loc}, \textit{guards})$ returns the fraction of countries
not containing \textit{guards} or on the virtual links between \textit{client\_loc} and
\textit{guards}.
\textsc{ExitSecurity}$(\textit{client\_loc}, \textit{dst\_loc}, \textit{guard}, \textit{exit})$
returns the fraction of countries either not containing \textit{guard} and not on the
$\{\textit{client\_loc}, guard\}$ virtual link or not containing \textit{exit} and not on the
$\{\textit{dst\_loc}, exit\}$ virtual link.

\subsection{Discussion} \label{subsec:algorithm:disc}
So far, we have been discussing trust-aware routing to destination IP
addresses. Many or most connections will require DNS resolution before
the destination IP is known. Exit relays resolve DNS requests in
current Tor routing to prevent linking of client IP address directly
to a destination DNS request. This must also be done in a trust-aware
manner, or there is little point in using trust-aware routing from exit
to destination once the IP address is known. If we rely on a chosen
exit to control the DNS resolution, then, even if it shares the default
trust values, it may not be a good exit for resolving the intended
destination.  When doing iterative DNS resolution from the client, possibly
switching to new circuits depending on the next identified DNS
resolver, the performance overhead could be significant. In this paper,
we assume that primary and backup nameserver ASes are
included with exit-relay descriptors. Assuming that these are
typically in the same AS as or immediately adjacent to the exit, this
will at least make sure that initial DNS requests are trust-aware.
How best to address DNS issues beyond that is outside the scope of this
paper.

We expect the \textsc{Cluster} process to be performed by organizations (\eg{}, Tor or EFF) and the
results distributed to users who trust their analysis and calculations.
End-users would then only need to perform the \textsc{Connect} process.
In the case of TrustAll, security depends on the assumption that
many other users are using the same policy.
Therefore, the TrustAll \textsc{Connect} process could be integrated with the standard Tor release
and enabled with a default policy (\eg{}, \textsf{The Man} or \textsf{Countries}). However, TrustOne
was designed to be used by a minority of users, and while the algorithm could be included with Tor,
it would not be enabled by default. We analyze the security and
performance of both approaches in the following sections.


%% file: sections/experiments.tex
\section{Security Analysis}\label{sec:sec}

\subsection{Experimental Setup}
We experimentally evaluate the security of the TrustAll and TrustOne algorithms against
\textsf{The Man} using an Internet map, data about the Tor network, and path-selection simulators
for \ps and for vanilla Tor. The AS-level routing map and past state of the Tor network are
constructed as described in Sec.~\ref{sec:attacks}, but for these experiments
we use data from December 2013.

We augment the routing map using sibling information based on RIPE WHOIS records. We identify IXPs
and place them on these on the AS-level paths using data from the IXP mapping
project~\cite{Augustin-IMC2009}. We group ASes into commercial  organizations using the results of
Cai \etal~\cite{Cai12b}. We conservatively group IXPs into organizations based on similarities in
their listings in the Packet Clearing House and PeeringDB (see \cite{ccs2013-usersrouted} for
details).

We simulate path selection on past Tor networks using the Tor Path Simulator (TorPS)~\cite{ccs2013-usersrouted}. TorPS provides Monte Carlo simulation of user circuit creation
over weeks and months on the changing Tor network. Each TorPS sample consists of a sequence of
circuits and assignments to those circuits of requested user connections over the period of
simulation. We use TorPS unmodified to evaluate the security of
vanilla Tor, and we also modify TorPS to use the \ps path selection algorithms.

We perform our TorPS simulations for two models of user behavior: the \emph{Typical} model, and
the \emph{IRC} model. {Johnson~\etal} describe these models in detail~\cite{ccs2013-usersrouted}.
The Typical model consists of four 20-minute user traces obtained from actual (volunteer) user
activity over Tor: (\emph{i}) Gmail / Google Chat, (\emph{ii}) Google Calendar / Docs, (\emph{iii})
Facebook, and (\emph{iv}) web search activity. It includes 205 unique destination IPs and uses TCP
ports 80 and 443.  These traces are played every day in five sessions between
9 a.m. and 6 p.m. This results in 2632 TCP connections per week. The IRC model uses the trace of a
single IRC session to \texttt{irc.oftc.net} on port 6697, which we observe to resolve to \texttt{82.195.75.116} in AS 8365 (TU Darmstadt, DE). This model repeatedly plays the trace 8 a.m.
to 5 p.m. every weekday. This results in 135 TCP connections per week. 

To evaluate security with respect to \textsf{The Man}, we use it to draw a sample of the compromised
relays and virtual links for each TorPS
sample and consider the security of that sampled path-selection behavior against that sampled
adversary. That is, we independently assign a compromised status to each AS organization, IXP
organization, and relay family using the probabilities given in Section~\ref{sec:metrics}. We then
consider the anonymity of the circuits in the TorPS sample against the sampled adversary.
We run simulations over the first week of December 2013. We use 3 guards for all simulations.

\subsection{Location Clusters}
The TrustAll algorithm prevents the chosen Tor paths from revealing client and destination 
locations beyond their location clusters. An adversary that can identify the relays in each position
of a circuit (\eg{}, by running a relay and being selected as a middle) may use them as evidence for
the clusters of the client and destination. For example, if
the adversary is also observing the exit-destination link, it may be the case that a given guard
and exit would only be used to visit that destination by members of a given client cluster.
As was shown in Section~\ref{sec:attacks}, this is an especially powerful attack if the
adversary can additionally link together multiple connections as
belonging to the same (pseudonymous) user.

Thus we must consider the anonymity that is afforded when a client or destination is known to
belong to a given cluster. In our experiments, we partition all Internet ASes into 200 clusters.
This number of clusters allows for significant diversity in cluster behavior while reducing the
anonymity set of roughly 3.7 billion addresses in IPv4 by a factor of 200. We perform clustering
using as the guard and exit locations the sets of ASes in which Tor guards and exits were observed
to reside during the six-month
period from June 2013 to November 2013, which precedes the simulation period in December 2013. 

Following the cluster-formation algorithm given in Section~\ref{sec:alg}, the 200 client clusters
are created by choosing as cluster representatives the top 200 Tor client ASes reported by
Juen~\cite{juen-masters}. In the resulting clustering, the median cluster size
in terms of contained addresses is 11,363,072, the minimum size is 10,840,321, and the maximum
size is 118,966,528.
Also as described in Section~\ref{sec:alg}, the destinations clusters were formed slightly
differently, using k-medoids clustering to identify representatives that were best able to
minimize distances between cluster members and their representatives.
The clusters that were the output of
this process had a median of 11,992,641 IPv4 addresses, with a minimum of 11,358,466 and a
maximum of 119,068,928. Our clustering algorithm sought to maximize the number of addresses
contained in each cluster, but it could easily incorporate other anonymity concerns, such as
AS or country diversity.

\subsection{TrustAll Security} \label{subsec:sec:trustall}
First we consider security against \textsf{The Man} when all users use \ps as the default
path-selection
algorithm (\ie{}, TrustAll). In particular we consider the threat of complete deanonymization via 
first-last correlation. We used the security parameters
$(\alpha^{su}_g, \alpha^{sc}_g, \alpha^{au}_g, \alpha^{ac}_g) = (0.95, 2.0, 0.5, 5.0)$,
$(\alpha^{su}_e, \alpha^{sc}_e, \alpha^{au}_e, \alpha^{ac}_e) = (0.95, 2.0, 0.1, 10.0)$, and
$(\alpha^w_g, \alpha^w_e) = (0.2, 0.2)$.
\begin{figure}[t]
\includegraphics[width=0.95\columnwidth]{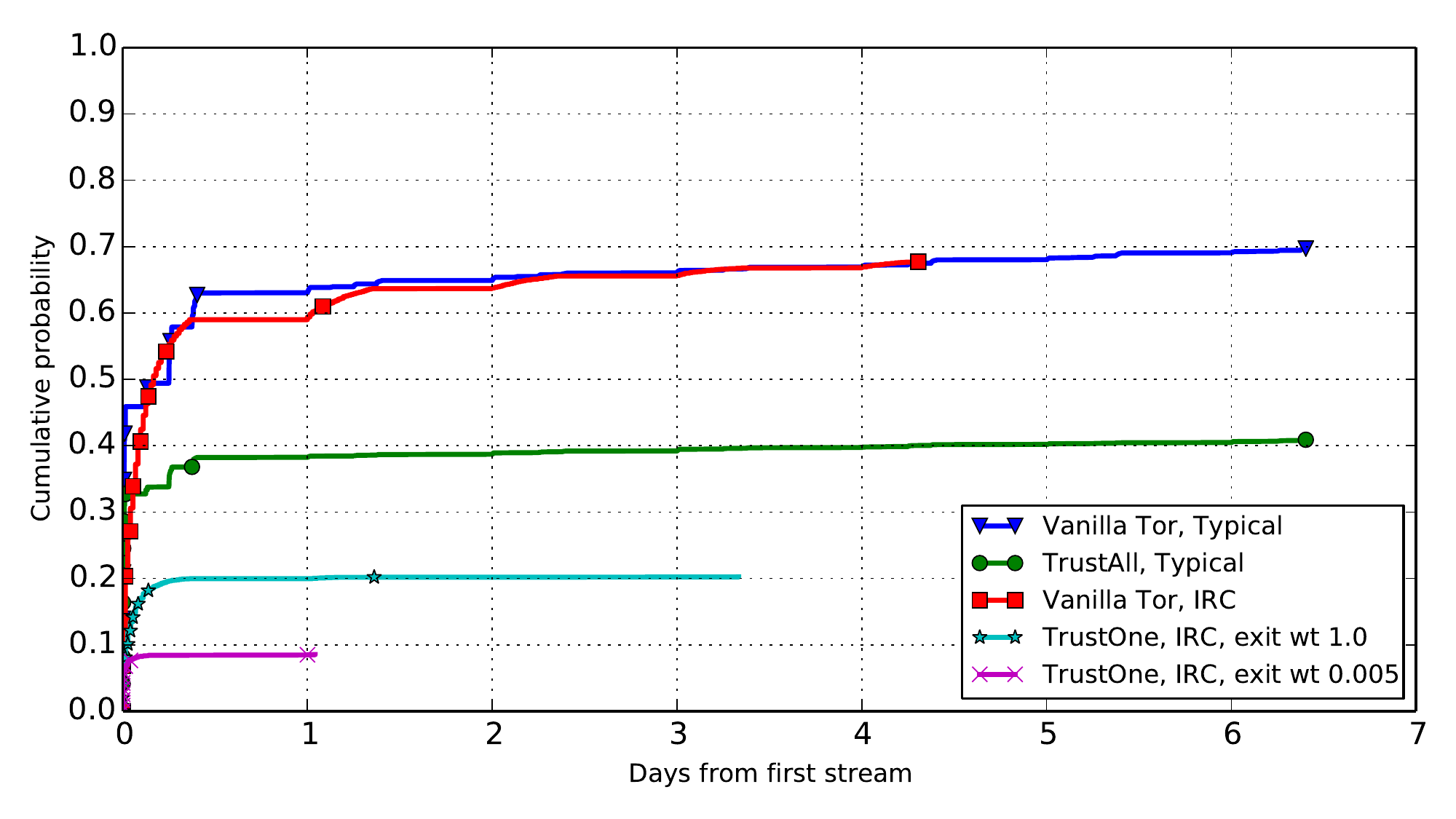}
\vspace{-4mm}
\caption{\small Time to first compromise in \textsf{The Man} model}
\vspace{-4mm}
\label{fig:the-man-time.cdf}
\end{figure}
\begin{figure}[t]
\includegraphics[width=0.95\columnwidth]{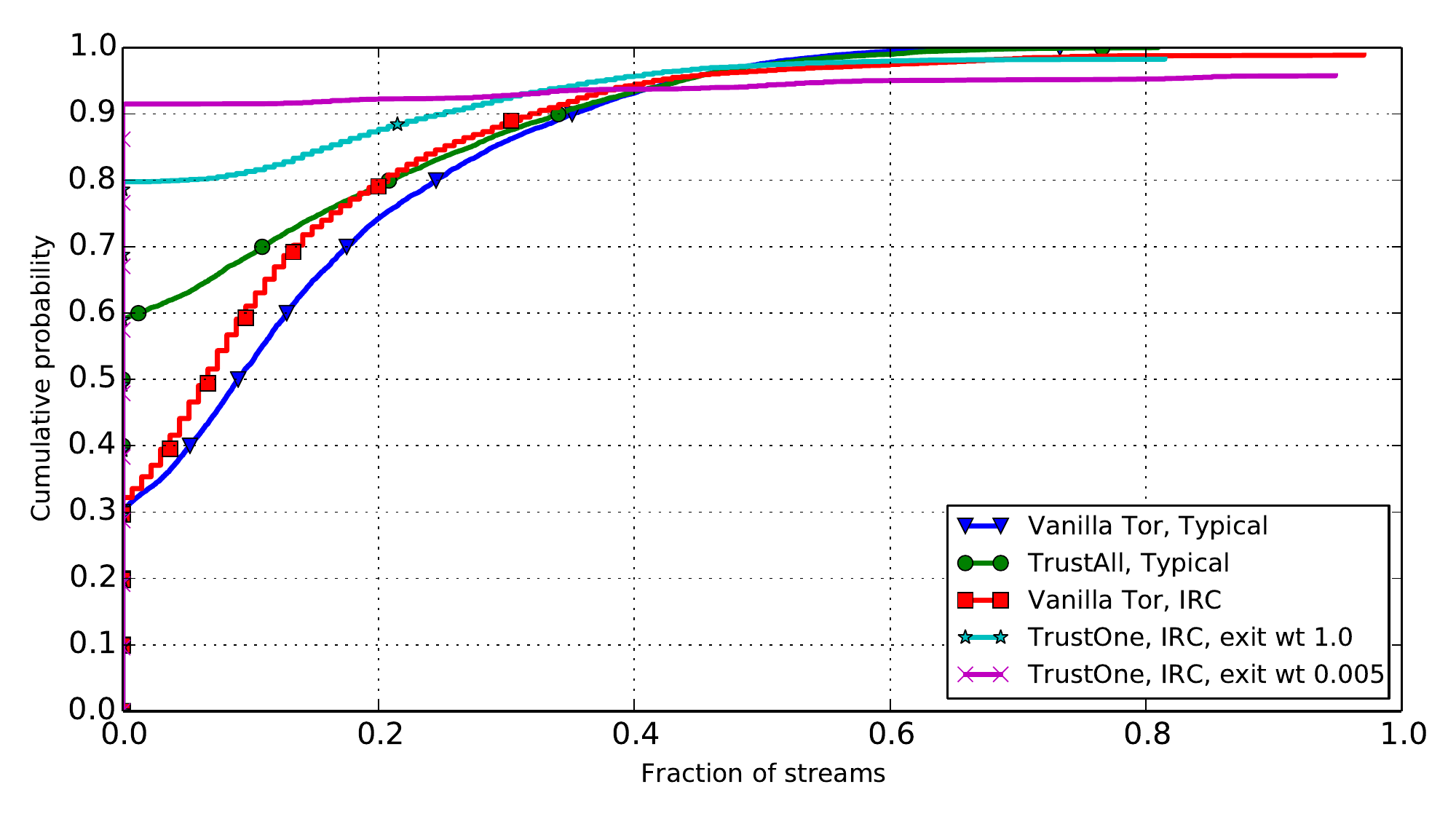}
\vspace{-4mm}
\caption{\small Fraction of compromised streams in \textsf{The Man} model}
\vspace{-4mm}
\label{fig:the-man-rate.cdf}
\end{figure}
Figures~\ref{fig:the-man-time.cdf} and~\ref{fig:the-man-rate.cdf} show cumulative distributions for
when and how often deanonymization occurs for a Typical user in the most popular Tor client AS
(6128) over 7 days of simulation.

We can see that TrustAll significantly reduces the chance of first-last correlation by
\textsf{The Man} as compared to vanilla Tor. Users coming from AS 6128 see the probability of at
least one successful first-last correlation attack drop from 0.7 to about 0.4. Observe that this
overall reduction occurs both because the chance of choosing a compromised guard among the initial
set of 3 is reduced (as seen in the values at 1 day) and because the chance of choosing additional 
compromised guards, precipitated by network churn, is reduced (as seen in the smaller slopes
of the CDF). The results also show that the median fraction of compromised streams drops from
around 0.1 to 0.

\begin{figure}[t]
\subfloat{\label{fig:countries-rate.cdf}{\includegraphics[width=0.95\columnwidth]{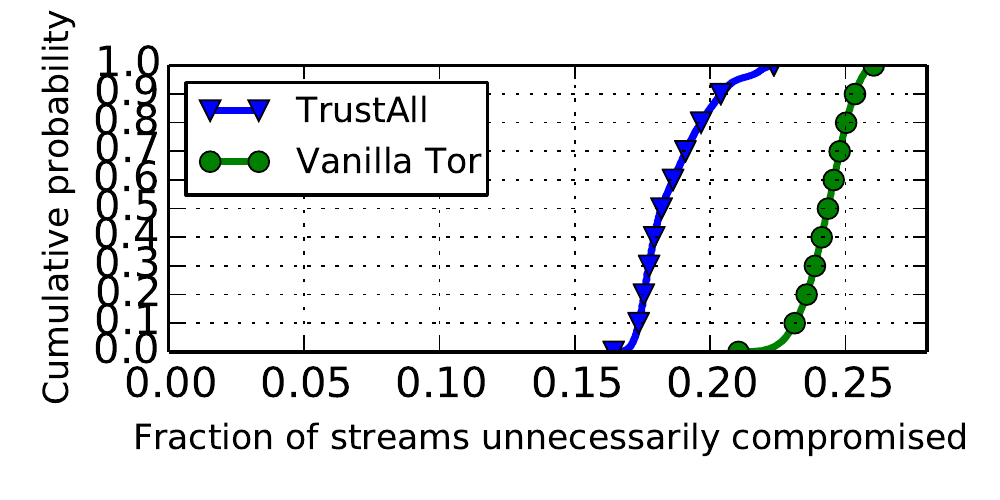}}}
\vspace{-4mm}
\caption{\small Security in the \textsf{Countries} model}
\vspace{-4mm}
\label{fig:countries}
\end{figure}

Next we consider the security of TrustAll in the \textsf{Countries} model. In this model, users
face multiple country adversaries (249), each of which deterministically compromises all ASes and
IXPs within its borders. In this setting, users are sometimes necessarily compromised against those
countries that contain both the source and destination AS. Thus we only consider the fraction of
those streams that are to a destination AS in different country than the client AS and are
``unnecessarily'' compromised by some country. Figure~\ref{fig:countries}
shows the distribution of this value for a Typical user in AS 6128 (which is in the US)
active over seven days. It shows that TrustAll reduces the fraction
of unnecessarily-compromised streams from a median of about 0.24 to a median of about 0.17.

Finally, we consider security of IRC users using the TrustOne algorithm when default users are
using vanilla Tor.
In this case, the \ps users choose guards and exits in a way dseigned to be sufficiently similar to
how they are selected in vanilla Tor. Specifically, guards are selected using
$\alpha^w_g = 0.005$ and
exits are selected using either $\alpha^w_e = 0.005$ or $\alpha^w_e = 1$. The former
weight for exits results in a TrustOne user being 200 times more likely to have chosen a given exit
than a vanilla Tor user. This could be an appropriate setting for a user who is not concerned with
revealing his use of TrustOne and his trust beliefs. It could also be
appropriate for a user who
is just trying to protect anonymity for a single connection considered in isolation. The weight
$\alpha^w_e = 1$ results in exit selection by a TrustOne user that is identical
to that of Tor users. This is an appropriate setting when the user wants to hide his use of
TrustOne and the user's adversaries may be able to link circuits together over time as belonging to
the same user.

Figures~\ref{fig:the-man-time.cdf} and~\ref{fig:the-man-rate.cdf} shows the chance of deanonymization of an IRC user in AS 6128 via first-last correlation
for TrustOne and vanilla Tor. We can see that TrustOne results in a significantly
lower chance of compromise, from a 0.68 chance for vanilla Tor users to about 0.2 or 0.1,
depending on the exit-selection parameter $\alpha^w_e$. The median compromise rate also drops
from about 0.7 to 0.

%% file: sections/performance.tex
\section{Performance Analysis}\label{sec:perf}

The \ps algorithm was designed to provide tunable
performance while improving users' security by limiting the probability that an
adversary can observe both sides of a Tor circuit. We now analyze the effect \ps
has on client performance and relay load balancing.


\subsection{Tor Network Model}

We evaluate the performance effects of \ps using
Shadow~\cite{jansen2012shadow,shadowweb}, a scalable and deterministic
discrete-event network simulator with a plug-in architecture that enables it to
run real software. 
Shadow runs the Tor software, and so we
can directly implement our algorithms in Tor's code base while increasing our
confidence that the application-level performance effects are realistic.

We configure a private Tor deployment using Shadow and the large-scale
topology produced by Jansen \etal~\cite{jansen2014kist}. Our base configuration
consists of 400 Tor relays (including 4 directory authorities and 93 exits), 1380 Tor
clients that also run a simple file-fetching application, and 500 simple file
servers. Of the clients, 1080 are \emph{Web} clients that are configured to: choose a
random server and download a 320 KiB file from it; pause for [1, 60] seconds
chosen uniformly at random; and repeat. 120 of the remaining clients are \emph{bulk}
clients that are configured to repeatedly download a 5 MiB file without pausing
between downloads. Each experiment is configured to run for 1 virtual hour, which
takes about 5 hours on our machine (using 12 Shadow worker threads) while
consuming 40 GiB of RAM.

We also run 180 \emph{ShadowPerf} clients and model their behavior after the
TorPerf~\cite{torperf} clients that measure and publish performance over time in
the public Tor network. We compared ShadowPerf to TorPerf performance over several
experiments and found that Shadow is able to model Tor performance characteristics
with reasonable accuracy over a range of file download sizes.


\subsection{\ps Implementation Details}

We branched Shadow~\cite{shadowgit} at commit \texttt{023055eb5},
shadow-plugin-tor~\cite{shadowtorgit} at commit \texttt{9eed6a7c5}, and Tor at
version \texttt{0.2.5.2-alpha} and modified them to support experimentation with
\ps. The implementation of both the TrustOne and TrustAll 
versions of \ps was done in 1126 lines of C code in Tor itself.


\begin{figure*}[t]
    \centering
    \subfloat[Time to first byte of download per client]{\label{fig:ttfb-trust-all-perf}{\includegraphics[width=0.33\textwidth]{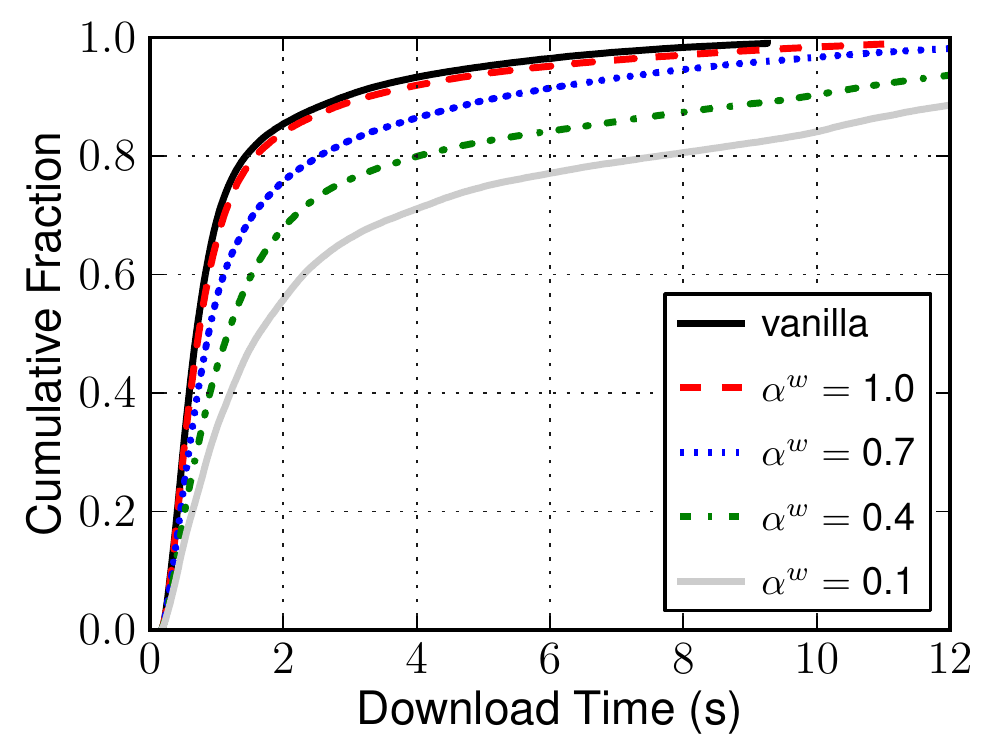}}}
    \subfloat[Time to last byte of 320KiB download per client]{\label{fig:ttlb-web-trust-all-perf}{\includegraphics[width=0.33\textwidth]{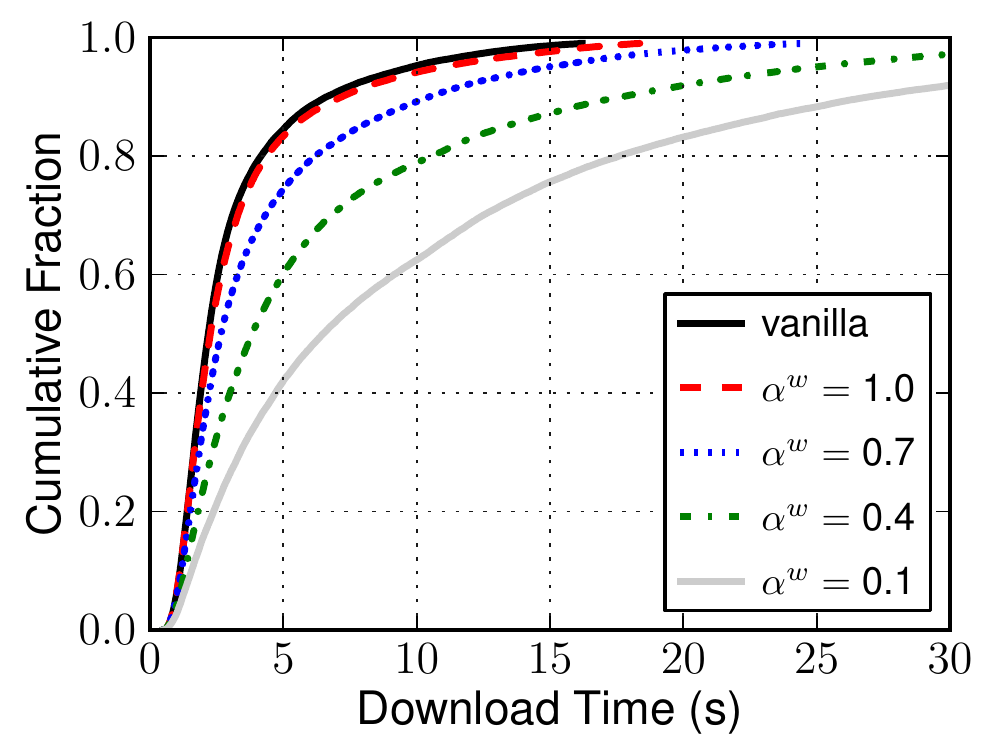}}}
    \subfloat[Time to last byte of 5MiB download per client]{\label{ttlb-bulk-trust-all-perf}{\includegraphics[width=0.33\textwidth]{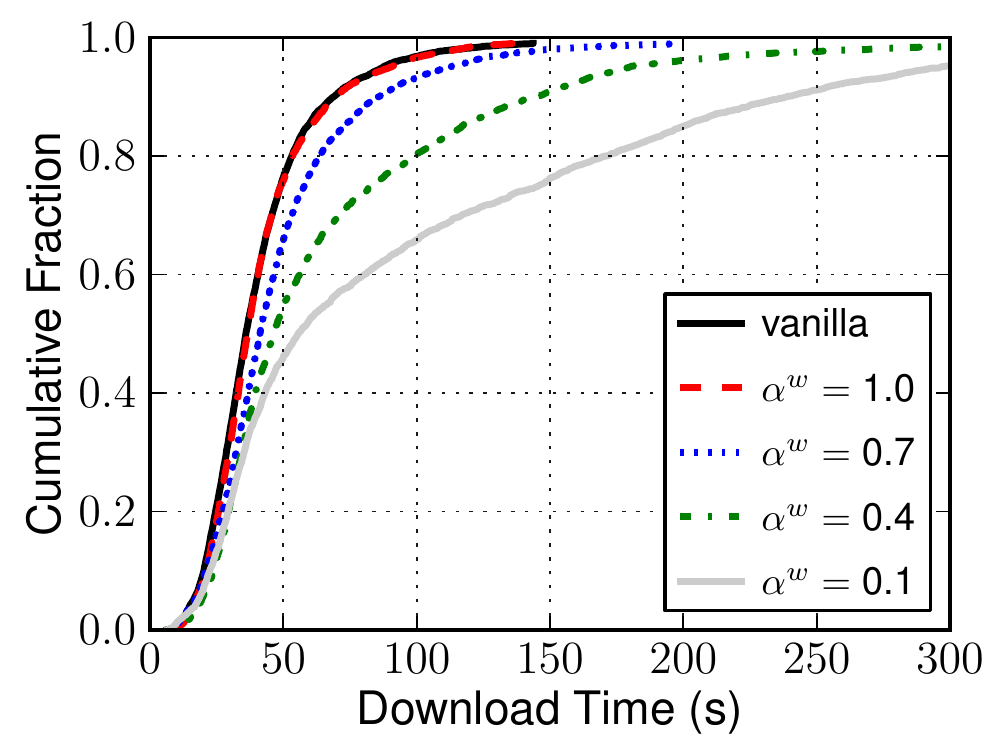}}}
    \\
	\subfloat[Number of downloads per client]{\label{fig:ndl-total-trust-all-perf}{\includegraphics[width=0.338\textwidth]{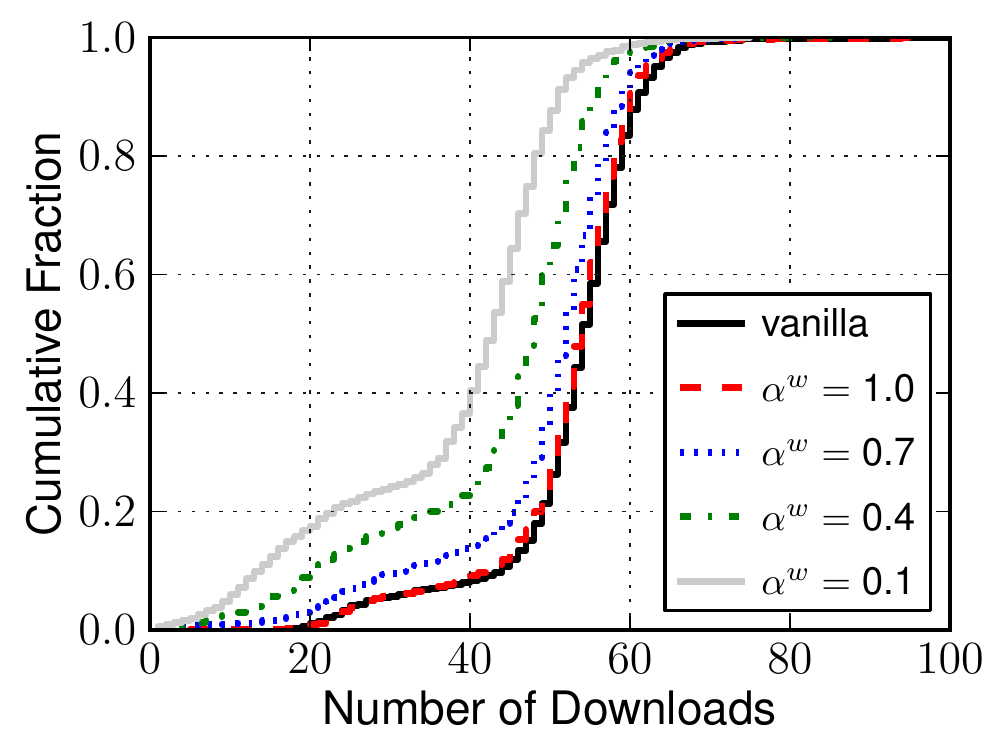}}}
	\subfloat[Aggregate relay throughput per second]{\label{fig:tor-bwcdf-trust-all-perf}{\includegraphics[width=0.33\textwidth]{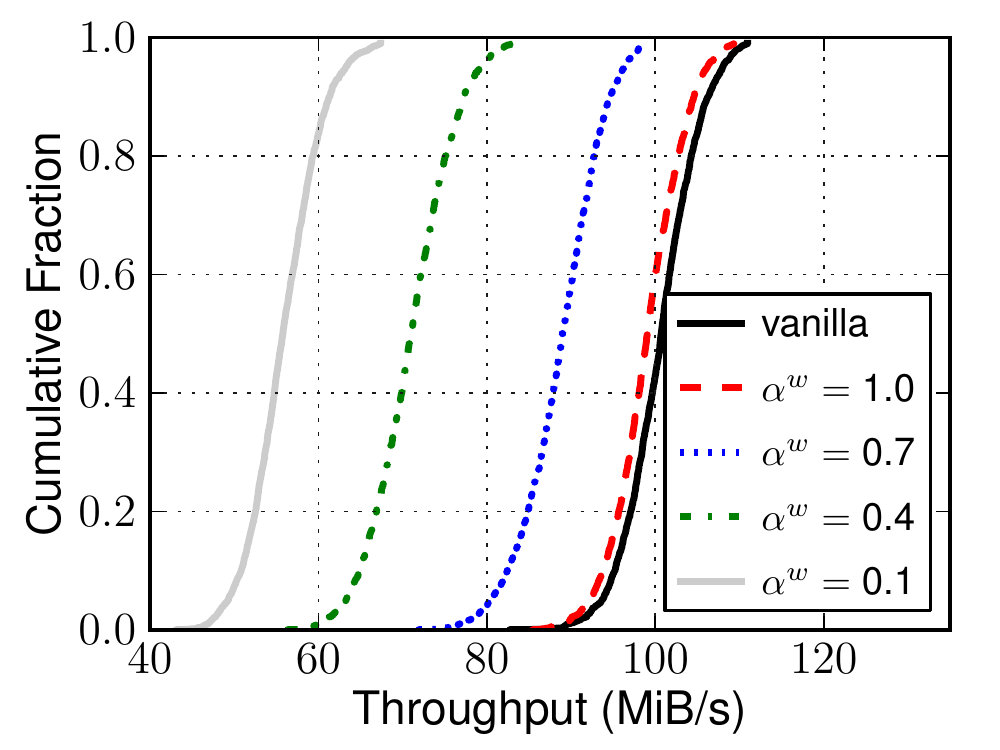}}}
	\subfloat[Unsafe consensus weights considered]{\label{fig:choices-trust-all-theman}{\includegraphics[width=0.33\textwidth]{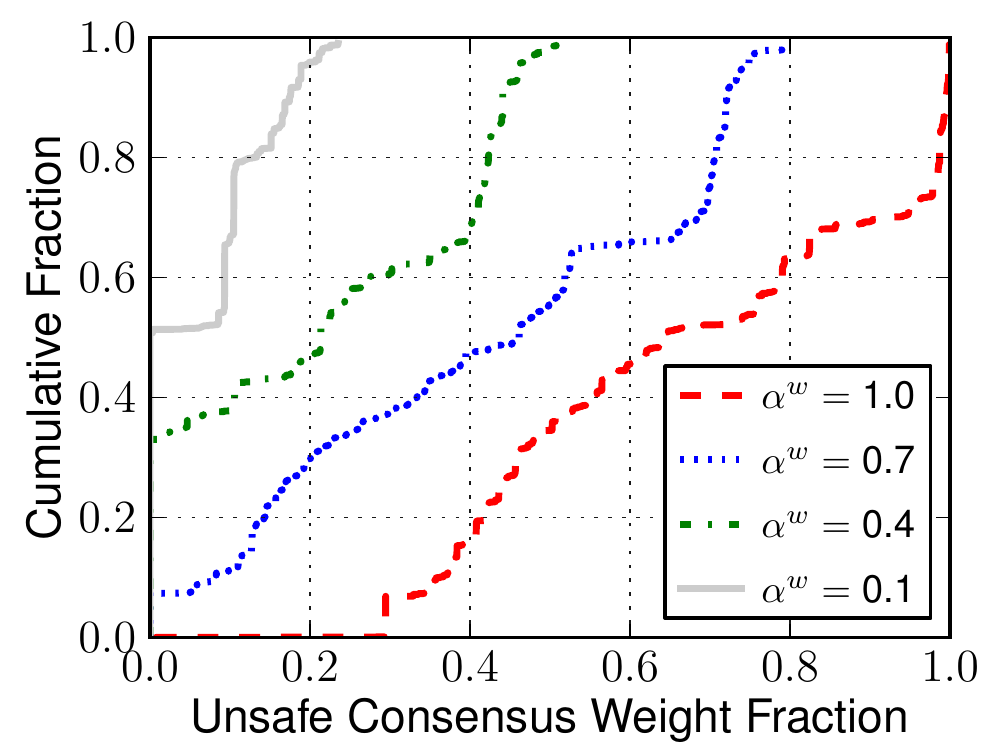}}}
    \caption{\small Performance of the TrustAll variation of \ps against \textsf{The Man} policy, varying required bandwidth fraction $\alpha^{w}$}
    \label{fig:perf-trust-all}
    \vspace{-4mm}
\end{figure*}


\subsection{TrustAll Against \textsf{The Man}}

Recall that in the TrustAll variation of \ps, all users in the network are
configured to select paths based on a common trust policy.  We explore the performance of \ps under different configurations of the parameters described in \S\ref{sec:alg}. We use the same values of the parameters defining safe and acceptable relays in position $p\in \{g, e\}$ (\ie{}, $\alpha^{su}_{p}$, $\alpha^{sc}_{p}$, $\alpha^{au}_{p}$, and $\alpha^{ac}_{p}$) that were used for
the security experiments in \S~\ref{subsec:sec:trustall}.
We then adjust the required bandwidth fraction $\alpha^w = \alpha^w_g = \alpha^w_e$ in order to
adjust the amount of load balancing that happens due to client path selection. Higher
values of $\alpha^w$ relax security by requiring clients to consider more
nodes in an attempt to exceed the bandwidth fraction and allow the algorithm
to better distribute load among relays that have the capacity to support it
(relays that are not safe or acceptable are never chosen in
any case). Lower values of $\alpha^w$ reduce the number of relays that a
client must consider, which means they effectively prefer more secure relays
and perform less load balancing. We experiment with different
values of $\alpha^w$ to explore these effects.

The results of our experiments are shown in Figure~\ref{fig:perf-trust-all}.
Figure~\ref{fig:ttfb-trust-all-perf} shows the distribution of the time to
receive the first byte of each download aggregated across all clients in our
network. As can be seen, there is a significant and consistent improvement in
latency to the first byte as $\alpha^w$ increases. As $\alpha^w$ increases,
client load is better distributed because more clients will end up choosing high
capacity nodes even if they are not the most secure choice. Similar results are
shown for time to complete the downloads, for Web clients in
Figure~\ref{fig:ttlb-web-trust-all-perf} and bulk clients in
Figure~\ref{ttlb-bulk-trust-all-perf}. Also,
performance differences are consistent across the $\alpha^w$ settings for both Web and
bulk clients, which we would expect because our path-selection algorithm is the same in
both cases.

Our experiments resulted in decreasing performance as $\alpha^w$ decreases. We expect this
to be the case since any deviation from Tor's default bandwidth-weighted algorithm
will result in suboptimal load balancing. However, our results indicate that a clear
performance-security trade-off is possible in \ps and that the algorithm
can be tuned to a desired level of performance while still removing the least
secure relays from consideration.

A side effect of the decrease in performance is fewer completed downloads by
each client over the course of the experiment due to our behavior models,
as evident in figure~\ref{fig:ndl-total-trust-all-perf}.
Related to download times, there is a significant reduction in the
number of downloads for clients (and a long neck for about 20 percent of Web
clients). This is likely due to the fact that these clients, because of their
location, consistently choose low capacity guards and exits that cause their
downloads to receive bad performance. (Clients in the long neck of number of
downloads are also in the long tail of download times.) This is also a result of
our behavior models, in which clients do not start a new download until the
previous one finishes. A richer behavior model in which some clients start multiple
downloads at a time (\eg{}, representing users opening multiple tabs or starting
multiple background file transfers) could alleviate this artifact.


As shown in Figure~\ref{fig:tor-bwcdf-trust-all-perf}, the reduction in the
number of downloads also reduces total aggregate network throughput (bytes
written summed across all relays every second). This again indicates a reduction
in the ability of Tor to properly balance load when all clients in the network use \ps.
Again, $\alpha^w=1.0$ performs the closest to vanilla Tor and does not result in a significant loss in performance, despite removing the least secure relays during path selection.

Finally, Figure~\ref{fig:choices-trust-all-theman} shows the cumulative fraction
of bandwidth weight from relays that fall outside of the safe thresholds but
that were still considered during path selection. These relays represent those
that were within the acceptable thresholds but not within the safe thresholds.
Recall that TrustAll selects relays in this acceptable zone one at a time, from most
to least secure, until the desired consensus weight fraction $\alpha^w$ is
reached. As expected, the more performance that is demanded (\ie{}, as
$\alpha^w$ increases), the more relays outside of the safe thresholds must be
used to reach the desired performance. Our results indicate that there are settings
of $\alpha^w$ that result in performance nearly as good as Tor's default
performance-optimized algorithm, while also taking security into consideration.

\subsection{TrustAll Against \textsf{Countries}}

The experimental results discussed above were obtained using \textsf{The Man} policy. For
completeness, we also experimented with the same parameters using the \textsf{Countries}
policy.
We confirmed that the same trends are present against the \textsf{Countries} policy as were
discussed above, and the results increased our confidence in the conclusions drawn about
the performance of \ps. (The full set of graphs are excluded for space reasons.)

\subsection{Trading Security for Performance}

\begin{figure}[t]
    \centering
    \subfloat[\textsf{The Man} Policy]{\label{fig:tradeoffs-theman}{\includegraphics[width=0.75\columnwidth]{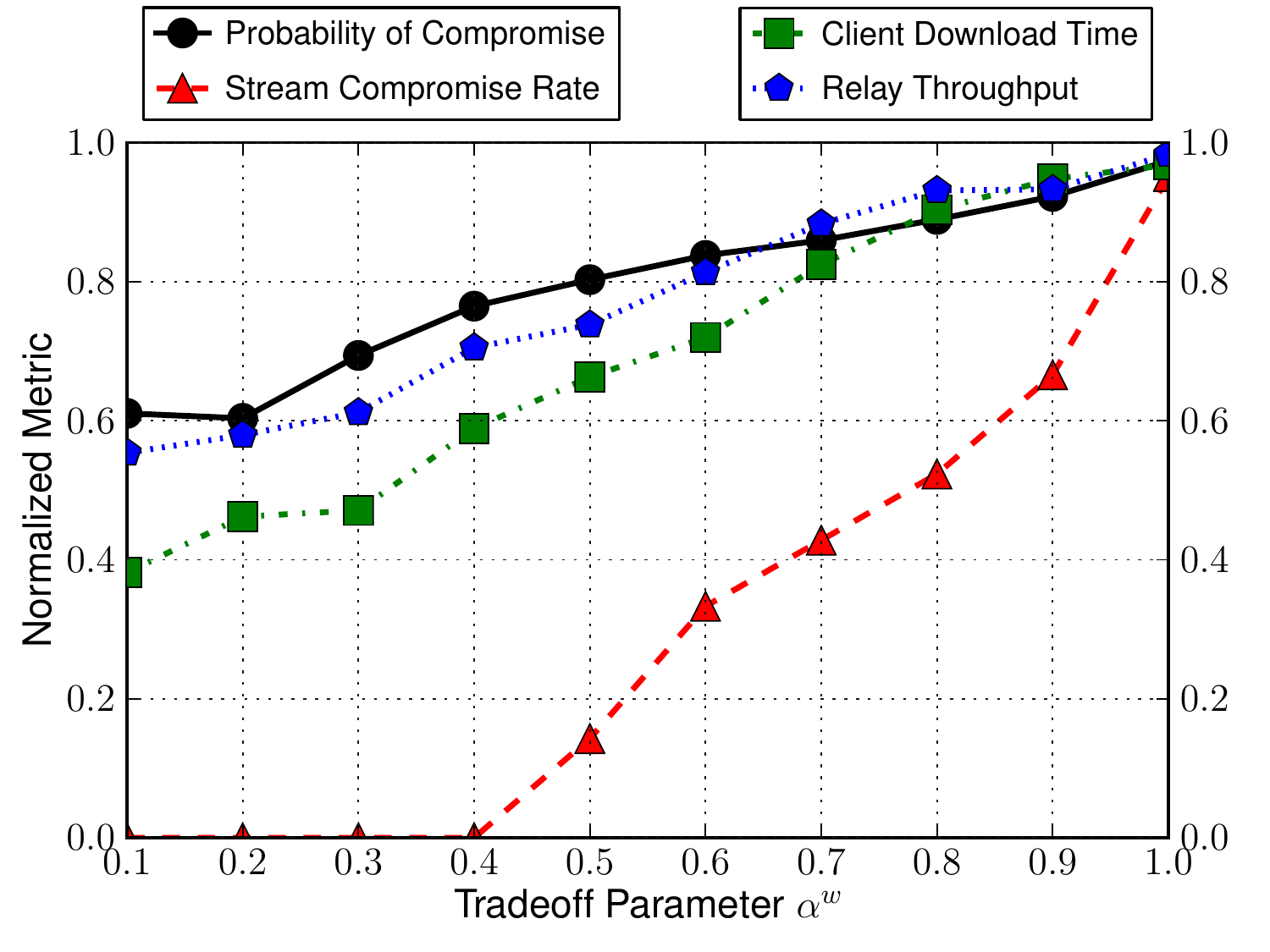}}}
    \vspace{-2mm}
    \\
    \subfloat[\textsf{Countries} Policy]{\label{fig:tradeoffs-country}{\includegraphics[width=0.75\columnwidth]{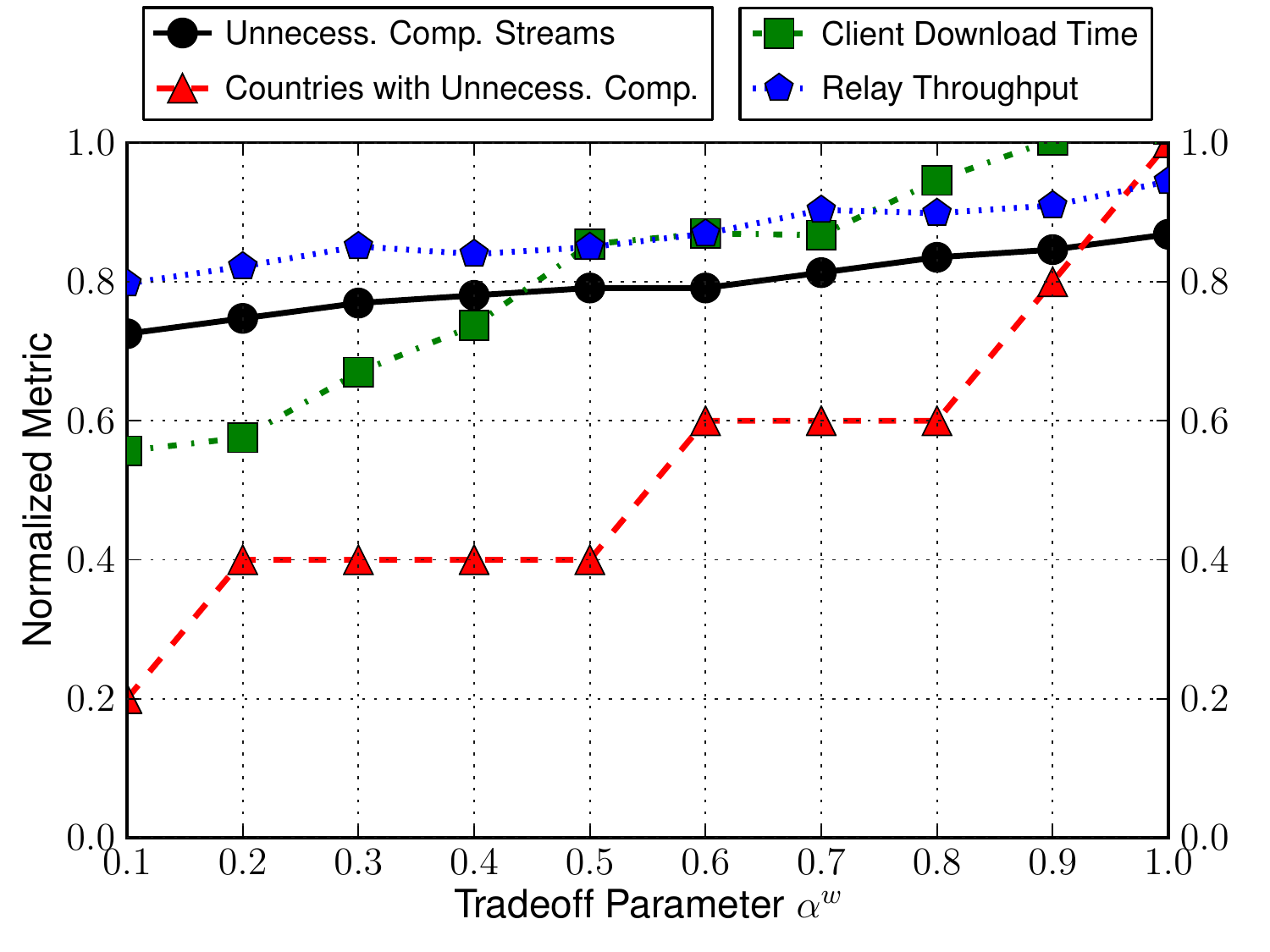}}}
    \vspace{-2mm}
    \caption{\small Trading performance and security in TrustAll with $\alpha^w$}
    \vspace{-4mm}
    \label{fig:tradeoffs}
\end{figure}

Figure~\ref{fig:tradeoffs} demonstrates how \ps directly trades performance for
security according to the parameter $\alpha^w$.
Figure~\ref{fig:tradeoffs-theman} shows the security-performance tradeoffs of
TrustAll against \textsf{The Man} policy for various values of $\alpha^w$. Shown in the
figure are two performance metrics: ``Client Download Time'' represents the
median across all clients of the median download time for each client; ``Relay
Throughput'' represents the median application throughput in terms of bytes
written per second, accross all relays over all seconds during the experiments.
Both of these metrics are normalized with respect to vanilla Tor, meaning that
values closer to 1.0 indicates that \ps achieves performance more similar to
that achieved by vanilla Tor. Also shown in Figure~\ref{fig:tradeoffs-theman}
are the ``Probability of Compromise'' and the ``Stream Compromise Rate'' as
metrics of security. The metrics are again normalized with respect to vanilla
Tor, so that values closer to 0 are less similar to vanilla Tor and indicate
higher security. As is clear in the figure, as the tradeoff parameter
$\alpha^w$ increases, both of the performance metrics improve while both of
the security metrics get worse. This is expected: as more relays are used to
reach the performance requirements of $\alpha^w$, it is more likely that
insecure relays or relays that exist on insecure paths will be selected and used
in a circuit.

A similar analysis applies to the \textsf{Countries} policy, the results for which are
shown in Figure~\ref{fig:tradeoffs-country}. The security metrics include the
median fraction of ``Unnecessarily Compromised Streams'', where the source and
destination of a stream do not exist in the same country and yet the stream was
still compromised, and the median number of countries with which the client
unnecessarily had a compromised circuit. The performance metrics are as above.
The same basic trends hold for the \textsf{Countries} policy: as $\alpha^w$
increases and the number of potentially unsafe relays considered for a path
increases, so does the number of avoidable stream compromises and the number of
countries to which a given client is unnecessarily compromised. In all
cases, however, security improves with respect to vanilla Tor while performance
decreases depending on the tunable setting of the tradeoff parameter
$\alpha^w$.

\subsection{TrustOne Against \textsf{The Man}}
In order for TrustAll to be effective, most clients must use it. If only a minority of clients
use trust, then they should use TrustOne in order to blend in with vanilla-Tor users. They can
also take advantage of their minority status by using higher-security parameters without
affecting Tor's load balancing much.


\begin{table}
\begin{center}
\begin{tabular}{|c||c|c|c|}\hline
metric & $\alpha^w=0.005$ & $\alpha^w=1.0$ & vanilla \\ \hline
\hline
Time to First Byte       & 0.870, 1.548   & 0.783, 1.694   & 0.690, 1.419 \\ \hline
Time to Last Byte 320KiB & 3.806, 3.785   & 2.685, 3.255   & 2.172, 2.597 \\ \hline
Time to Last Byte 5MiB   & 39.825, 29.342 & 35.203, 14.395 & 35.777, 20.658 \\ \hline
Tor Throughput (MiB/s)   & 98.635, 4.893  & 99.699, 5.387  & 100.660, 4.250 \\ \hline
\end{tabular}
\end{center}
\caption{\small Statistical summary (median, standard deviation) of performance for TrustOne}\label{tab:perf-trust-one}
\vspace{-4mm}
\end{table}

We demonstrate the performance of TrustOne by configuring 68 of our Web clients
and 5 of our bulk clients to run the
TrustOne algorithm with $\alpha^w_g = 0.005$ and
$\alpha^w_e\in \{0.005,1.0\}$; the other parameter settings are as in the
TrustAll experiments. All other clients use the vanilla-Tor path-selection
algorithm. Thus the TrustOne clients choose secure guards, and they either choose exits
identically to vanilla-Tor users in order to blend in ($\alpha^w_e=1.0$)
or don't attempt to hide their use of TrustOne and instead choose exits very securely
($\alpha^w_e=0.005$).


Table~\ref{tab:perf-trust-one} provides a statistical summary of the performance of 
the vanilla and trust clients. Note that
the results for $\alpha^w\in \{0.005, 1.0\}$ come from two separate TrustOne experiments,
the vanilla-Tor results come from another experiment with no TrustOne clients, and
the reported download times for $\alpha^w\in \{0.005, 1.0\}$ are only for TrustOne clients.
Across all three client performance metrics (time to first byte, and
time to last byte of Web and bulk downloads), we see only a small drop
in client performance for both settings tested. Although our sample size is
small, both settings of $\alpha^w_e$ resulted in similar performance for the
trusted user set. This indicates that performance for those clients was due to
the capacity and congestion of their guard nodes (which they chose using a
secure value of $\alpha^w_g$). Also shown in Table~\ref{tab:perf-trust-one} are
results showing that relay throughput in the TrustOne experiments
was not significantly lower than in the vanilla Tor experiment
(relay throughput is over all relays and thus in the TrustOne experiments includes
traffic from both trust-aware and vanilla clients).
This is attributable to a relatively small change in the load balancing across
the network since only the trust users deviate from the optimized load balancing
algorithm. Our results indicate that little performance is lost from using the
TrustOne algorithm when a relatively small set of users are doing so.

%% file: sections/errors.tex

%
%

\section{Trust Errors}\label{sec:errors}

Because the client's paths depend on her beliefs and may not be accurate, it is important to investigate the effects of errors in the client's beliefs on her security.  Here, we do that experimentally by considering a variety of mismatches between client trust beliefs and the actual adversary.  We look at three client behaviors against nine different actual adversaries for a single location (AS 6128) over one week.  We also look at our Typical client in 401 different locations (the client ASes observed by Juen~\cite{juen-masters} and in our AS-level map for December 2013) with trust beliefs corresponding to \textsf{The Man}, but where the actual adversary distribution is one of a selected set of other behaviors.  

The client might also have beliefs about the structure of the network.  Errors in those may have significant impacts on the client's security if, for example, the client believes that an untrustworthy AS organization does not contain an AS that it actually does. We focus our experiments here on errors in trust beliefs, however. 

We consider three different client behaviors:  The Typical and IRC clients with \textsf{The Man}
policy are as described above.  We also consider a client with Typical network behavior who chooses
paths based on trust beliefs that match the \textsf{Countries} adversary.  

These client properties are combined with various adversaries, which may or may not match the client's beliefs and policies:\\
    \textit{Type 0:} The adversary is \textsf{The Man} adversary described above.\\
	\textit{Type 1:} The probability of compromise is increased, relative to \textsf{The Man}, by a factor of $1.25$ for AS/IXP organizations, lone ASes, and relay families.  This reflects the possibility that the client uniformly underestimates the adversary's capabilities.\\
	\textit{Type 2a:} This is the same as \textsf{The Man} except ASes that are not part of any organization are not compromised.\\
	\textit{Type 2b:} This is the same as \textsf{The Man} except ASes that are not part of any organization are compromised with probability $0.05$.\\
	\textit{Type 3:} For each run, half of the AS organizations and half of the IXP organizations are compromised with probability $0.15$, and the others are compromised with probability $0.05$. For
efficiency, an AS that is not part of an AS organization is only assigned a compromised status
when it is first encountered on a virtual link during analysis. Upon its initial observation, such
an AS is assigned one of $0.15$ and $0.05$ as a compromise probability using a fair coin, and then
it is compromised with that probability.\\
	\textit{Type 4:} The adversary is the same as \textsf{The Man} except longer uptime \emph{increases} the compromise probability for a relay family (\eg, because of the increased chance of unpatched software).  In particular, the compromise probability for a relay family with uptime $t_f$ is $0.1 - (0.1-0.02)/(t_f + 1)$.\\
	\textit{Type 5:} The adversary compromises each relay with probability $0.1$ and each virtual link with probability $0.3439 = 1-0.9^4$.  (The latter value is chosen to approximate the probability of compromising ASes/IXPs independently.  On the virtual links that we consider, the median number of ASes/IXPs is four, although these are not necessarily from distinct organizations.)\\
	\textit{Type 6:} The adversary is the same as \textsf{The Man} for ASes/IXPs.  For relays and relay families, the adversary compromises nontrivial families with probability $0.1$ and individual relays that are not part of a nontrivial family is $0.05$.\\
	\textit{Type 7:} The adversary is the same as \textsf{The Man} for ASes/IXPs.  For relays and relay families, the adversary compromises families with probability $p_{max} - (p_{max} - p_{min}) 2^{-(f_{size} - 1)}$, where $p_{min}$ and $p_{max}$ are the minimum ($0.02$) and maximum ($0.1$) probabilities of family compromise for \textsf{The Man} and $f_{size}$ is the number of relays in the family.

Table~\ref{tab:weeklong-errors} shows the median time to first compromise (TTFC) in days and the probability that some circuit is compromised for the three different client types and nine different adversary distributions noted above.  In each case, we take the client to be in AS 6128.  The data are obtained from 10,000 simulations of client behavior from 12/1/13 to 12/7/13.  Values of ``$>7$'' for the TTFC indicate that the value is at least seven days, the length of these experiments.

\begin{table}
\begin{center}
\begin{tabular}{|l|c|c|c|c|c|c|c|c|c|}\hline
\multicolumn{10}{|c|}{Client: Typical, against \textsf{The Man}}\\ \hline
Adv.$\rightarrow$ & 0 & 1 & 2a & 2b & 3 & 4 & 5 & 6 & 7 \\ \hline
TTFC & $>7$ & $>7$ & $>7$ & $>7$ & $>7$ & $>7$ & $.01$ & $>7$ &  $>7$  \\ \hline
Prob. & $.4$ & $.49$ & $.4$ & $.4$ & $.41$ & $.47$ & $.68$ & $.47$ &  $.47$  \\ \hline\hline
\multicolumn{10}{|c|}{Client: IRC, against \textsf{The Man}}\\ \hline
Adv.$\rightarrow$ & 0 & 1 & 2a & 2b & 3 & 4 & 5 & 6 & 7 \\ \hline
TTFC & $>7$ & $4.17$ & $>7$ & $>7$ & $>7$ & $>7$ & $.07$ & $>7$ & $>7$  \\ \hline
Prob. & $.41$ & $.5$ & $.41$ & $.41$ & $.41$ & $.48$ & $.71$ & $.49$ &  $.48$  \\ \hline\hline
\multicolumn{10}{|c|}{Client: Typical, against \textsf{Countries}}\\ \hline
Adv.$\rightarrow$ & 0 & 1 & 2a & 2b & 3 & 4 & 5 & 6 & 7 \\ \hline
TTFC & $.38$ & $.01$ & $.38$ & $.38$ & $.38$ & $.26$ & $.0$ & $.25$ &  $.38$  \\ \hline
Prob. & $.58$ & $.66$ & $.56$ & $.57$ & $.57$ & $.6$ & $.82$ & $.61$ &  $.58$  \\ 
\hline
\end{tabular}
\end{center}
\vspace{-2mm}
\caption{\small Median time to first compromise (in days) and probability of compromise for three different client behaviors and nine different actual adversary distributions.  Data are from 10,000 simulations of a client in AS 6128 running for 7 days.}\label{tab:weeklong-errors}
\vspace{-4mm}
\end{table}

Table~\ref{tab:day-errors} shows various compromise statistics for a Typical client who chooses paths based on beliefs that match \textsf{The Man} against three different adversary distributions.  For each of the 401 client locations, we ran 10,000 simulations and took the median TTFC, compromise probability, and fraction of compromised paths for that location.  The table shows the minimum, median, and maximum of these per-location median values.  Values of ``$>1$'' for the TTFC indicate that the value is at least one day, the length of these experiments.

\begin{table}
\begin{center}
\begin{tabular}{|l|c|c|c|c|c|c|c|c|c|}\hline
Adv. & \multicolumn{3}{|c|}{Med.\ TTFC} & \multicolumn{3}{|c|}{Med.\ Prob.} & \multicolumn{3}{|c|}{Med.\ Frac.} \\ \hline
 & \multicolumn{3}{|c|}{min./med./max.} & \multicolumn{3}{|c|}{min./med./max.} & \multicolumn{3}{|c|}{min./med./max.} \\ \hline
2a & $.01$ & $>1$ & $>1$ & $.21$ & $.45$ & $.66$ & $.0$ & $.0$ &  $.09$  \\ \hline
4 & $.01$ & $>1$ & $>1$ & $.27$ & $.49$ & $.69$ & $.0$ & $.0$ &  $.11$  \\ \hline
5 & $.0$ & $.01$ & $.01$ & $.59$ & $.66$ & $.79$ & $.1$ & $.13$ &  $.16$  \\ \hline
\end{tabular}
\end{center}
\vspace{-2mm}
\caption{\small Statistics (minimum, median, and maximum) on the median times to first compromise (in days), compromise probability, and fraction of paths compromised over 10,000 trials for each of 401 client locations running for one day each against a selected set of adversaries.  The clients choose paths against \textsf{The Man}; the actual adversary is shown in the first column.}\label{tab:day-errors}
\vspace{-4mm}
\end{table}

Comparing Table~\ref{tab:weeklong-errors} with Fig.~\ref{fig:the-man-time.cdf}, we see that when the
client in AS 6128 chooses paths against \textsf{The Man}, the use of \ps increases her security,
compared with vanilla Tor, against adversaries that are related to \textsf{The Man} even though the
client is wrong about the exact nature of the adversary. More precisely, this is true for all of
the adversary types considered in this section other than Type 5, which is the adversary that
independently compromises each relay and virtual link and is the only type that does not
compromise the network based on organizations and families. When the adversary is actually of Type
5, Tables~\ref{tab:weeklong-errors} and \ref{tab:day-errors} show that it is able to do quite well
against the client over many locations and client behaviors.

%% file: sections/propagation.tex
\section{Obtaining and Propagating Trust}\label{sec:propagation}

We consider as an example how much data must be stored and communicated
to implement \textsf{The Man} policy.
First, the client must be able to determine the cluster of itself and its
destinations. With 46368 ASes in the network map used for \ps analysis,
200 client clusters, and 200 destination
clusters, 182 KiB suffice for each client to determine the needed clusters.
Second, to choose guards and exits, the client needs the ability to
determine the AS and IXP organizations on any virtual link either
between their cluster representative and a guard or between
a destination-cluster representative and an exit. There are only 359
IXPs, and so an AS or IXP organizations can be specified in two bytes.
For data gathered December 2013, all guards are within 603 ASes, all
exits are within 962 ASes, and the average number of AS and IXP
organizations on a virtual link is 4.05. Thus a
list of the entities on all the relevant virtual links for a given
client is would be 1.68 MiB.
Routing changes could be propagated daily or weekly with much smaller
updates once the full data is obtained.




%% file: sections/related.tex
\section{Related Work}
\label{sec:related}
An early proposals to use trust for Tor routing
came from {\O}verlier and
Syverson~\cite{hs-attack06}, who suggest choosing guards
``based on trust in the node administrator''. However, they do not
develop this idea.
A mathematical notion of trust in Tor was introduced by Johnson and
Syverson~\cite{trusted-set}. They formalize trust as the probability
of compromise of a relay, and they provide an analysis
of end-to-end correlation attacks when there are just two different
levels of trust. This model was later used by Johnson
\etal{}~\cite{jsdm11ccs} to produce a ``downhill'' Tor path-selection
algorithm that can handle arbitrary trust levels at the relays and
is designed to prevent traffic-correlation attacks.
Jaggard \etal{}~\cite{trustrep-popets14} greatly expand this
probabilistic notion of trust by describing how to identify
compromise factors that can apply to the links as well as the nodes,
such as AS organizations, legal jurisdictions, and router software.
They focus on expressing such rich trust models,
while in this paper we focus on using these models in a path-selection
algorithm that improves security.

Another approach to trust for anonymous communication is to explicitly
leverage social network relations. Danezis \etal~\cite{drac-pet2010}
describe this for interactive but low-volume anonymous communication.
Concentrating on low-volume applications allowed them to make use of
padding, which is generally too expensive and too ineffective for some
of the more popular applications that use Tor. 
Mittal \etal~\cite{pisces-ndss13} describe a social-network onion-routing
architecture designed for Web browsing and other interactive
communications that adds resistance to an active
adversary. This design uses
potentially much longer paths than Tor's three hops to achieve
intended security, and performance may suffer significantly as a
result of this and other features of the design.

The threat of AS adversaries to Tor was first recognized
by Feamster and Dingledine~\cite{feamster:wpes2004}. Their analysis
shows that entry and exit paths through the network are likely to be
simultaneously observed by a single AS 10\% to 30\% of the time, depending
on the locations of the client and destination. They
suggest that clients choose entry and exit nodes to avoid traversing the same
AS upon entry and exit.
Edman and Syverson~\cite{tor-as} update this work and show that,
despite the growth of the network from
about 30 to about 1300 relays, the risk of denanonymization by a
single AS is not reduced. They also show how to efficiently implement
the AS-aware path selection suggested by Feamster and Dingledine
by providing clients with routing data that enables them to infer
AS-level routing paths.
Murdoch and Zieli\'nski~\cite{murdoch:pet2007} introduce IXPs
as a potential adversary.
They show that an IXP
can correlate traffic even at low rates of sampling.
Link adversaries at both ASes and IXPs were extended by
Johnson \etal~\cite{ccs2013-usersrouted}
to consider adversaries controlling multiple ASes or IXPs, such as
companies that own many IXPs.
Akhoondi \etal~\cite{lastor} present an alternate method for clients to
efficiently infer the ASes between hosts for purposes of choosing Tor
paths that avoid allowing the same AS to observe entry and exit traffic.
Juen~\cite{juen-masters} presents another method for this purpose,
this time with the addition of inferring IXPs on those paths. All of
the preceding suggestions for AS-aware tor path selection neglect
key details, such as how circuits are reused and how
to handle destinations with no path that avoids putting an AS on both
sides. In addition, Juen \etal~\cite{tortraceroutes-pets15} show that
methods of AS inference for detecting Tor paths vulnerable to AS-level
compromise suffer from significant false-positives and false-negatives
when compared to direct traceroute measurements.

Nithyanand \etal~\cite{astoria-ndss2016} present Astoria, which is
the first reasonably-complete network-aware Tor path-selection algorithm.
As described in Section~\ref{sec:attacks}, like other previous work on
network-aware path selection, Astoria is only secure when each connection
is analyzed independently. DeNASA~\cite{denasa-pets2016}, by Barton and
Wright, is another recent and fully-specified
network-aware Tor path-selection algorithm. DeNASA only considers as
adversaries individual ASes, and chooses to just protect against the eight
ASes that are most likely to be in a position to deanonymize a connection.
DeNASA also doesn't consider the specific destination when constructing
a circuit, which allows it to use pre-built circuits for speed but makes it
unable to protect connections to destinations with paths dissimilar from
the pre-selected set used for exit selection. However, DeNASA is still
vulnerable to leakage about a client's AS across repeated connections
(assuming its guard and exit-selection algorithms are jointly used).

Sun \etal~\cite{raptor-usenix2015} show that traffic correlation attacks
on Tor are effective even when the attacker observes paths in different
directions on the entry and exit sides. They also demonstrate the application
of BGP hijacking and interception attacks to redirect
Tor traffic to malicious ASes in order to deanonymize users.
Tan \etal~\cite{tor-dataplane-defenses} extend this analysis and show that
90\% of Tor's bandwidth is vulnerable to BGP hijacking, and they propose
as a defense a set of monitors to detect routing attacks and notify Tor
clients to avoid the affected relays.


%% file: sections/conclusion.tex
\section{Conclusion}

In this paper, we show how previous network-aware Tor
path-selection algorithms are vulnerable to attacks across multiple Tor connections.
We present \ps, a path-selection algorithm
for Tor that is not vulnerable to such attacks and that enables clients to avoid
traffic-correlation attacks by using trust that they have in network elements.
We present two global-adversary models, analyze the security
and performance of \ps against these adversaries, and consider both trust errors and trust
propagation.